\documentclass[%
 amsmath,amssymb,
 reprint,%
]{revtex4-2}

\usepackage{graphicx}% Include Fig. files
\usepackage{dcolumn}% Align table columns on decimal point
\usepackage{bm}% bold math
\usepackage{stackengine}
\usepackage{overpic}
\usepackage{rotating}

\usepackage[utf8]{inputenc}
\usepackage[T1]{fontenc}
\usepackage{mathptmx}
\usepackage{etoolbox}
\usepackage{amsmath}

\DeclareMathOperator*{\argmin}{arg\,min}

\def\dagger{%
  \stackon[-0.25ex]{\rule{0.4pt}{1.0ex}}{\rule{.75ex}{0.4pt}}}

\makeatletter
\def\@email#1#2{%
 \endgroup
 \patchcmd{\titleblock@produce}
  {\frontmatter@RRAPformat}
  {\frontmatter@RRAPformat{\produce@RRAP{*#1\href{mailto:#2}{#2}}}\frontmatter@RRAPformat}
  {}{}
}%
\makeatother
\begin{document}
\bibliographystyle{unsrt}
\preprint{AIP/123-QED}

%Add Captions for all of the Fig.s
%Also give a description for how Travis is actually enforcing that the learned DMD models are stable

\title[]{Data-driven methods to discover stable linear models of the inductive helicity injectors}
% Force line breaks with \\
\author{Zachary Daniel}

\affiliation{Applied Physics \& Applied Mathematics, Columbia University, New York, NY 10027, USA }%

\author{Alan Kaptanoglu}%
\affiliation{Courant Institute of Mathematical Sciences, New York University, New York, NY 10012, USA}%

\author{Chris Hansen}
%\homepage{http://www.Second.institution.edu/~Charlie.Author.}
\affiliation{Applied Physics \& Applied Mathematics, Columbia University, New York, NY 10027, USA}%

\author{Kyle Morgan}
%\homepage{http://www.Second.institution.edu/~Charlie.Author.}
\affiliation{Zap Energy, Everett, WA 98203, USA}%

\author{Steven L. Brunton}
%\homepage{http://www.Second.institution.edu/~Charlie.Author.}
\affiliation{Mechanical Engineering, University of Washington, Seattle, WA 98195, USA}%

\author{J. Nathan Kutz}
%\homepage{http://www.Second.institution.edu/~Charlie.Author.}
\affiliation{Applied Mathematics, University of Washington, Seattle, WA 98195, USA}%
\affiliation{Electrical and Computer Engineering, University of Washington, Seattle, WA 98195, USA}

\date{\today}% It is always \today, today,

%Abstract outline:
%Tell the story:

\begin{abstract}
Accurate and efficient circuit models are necessary to control the power electronic circuits employed in most plasma physics experiments. Controlling the behavior of these circuits is inextricably linked to generating the desired plasma conditions. Linear models are greatly preferred for control applications due to their well-established performance guarantees, however, it is challenging to identify a linear approximation of the coupled nonlinear coil-plasma system from limited noisy measurements. In this work, the {\em Bagging Optimized Dynamic Mode Decomposition} (BOP-DMD) is leveraged to learn stable, reduced order models of the interaction between the coils and the plasma in the {\em Helicity Injected Torus -- Steady Inductive Upgrade} (HIT-SIU) experiment. BOP-DMD is trained and evaluated on an analytic model of the vacuum dynamics of the injector circuits, as well as an analytic linear reduced-order model for the circuit dynamics when a plasma is present.  BOP-DMD is then fit on experimental data, both on experiments with and without a plasma. In doing so, we demonstrate the capability of BOP-DMD to produce stable, linear models for control and uncertainty quantification in a high-power, coupled plasma-coil system.

\end{abstract}

\maketitle

\section{\label{sec:level1}Introduction:}
Forming and sustaining a plasma for experimental or applied purposes generally requires the use of applied electric/magnetic fields to ionize/heat the plasma, drive currents, and/or maintain or modify embedded magnetic fields. This creates a coupled plasma-circuit system that can introduce significant nonlinearity and additional couplings between circuit elements that are not present in vacuum. Accurate and robust control of such systems is required to achieve the desired plasma conditions, and to ensure that the device remains in safe operational limits. The problem of controlling the multi-scale and dynamic plasma formed in experimental and industrial devices remains an active area of research~\cite{gribov2007, ahedo2013helicon, seo2024avoiding, nelson2023, battey2023, fujimoto2002control}. Moreover, the models and algorithms employed in real-time control should be interpretable, and have quantifiable uncertainty. Many modern control algorithms require a model of the dynamics of interest~\cite{anirudh20232022},  or rely on a black-box, model-free control strategy~\cite{degrave2022magnetic}. Moreover, while there are a variety of models for describing the evolution of a plasma, many of the parameters upon which these equations depend are difficult to measure in modern experiments, or exhibit highly nonlinear behavior, making them difficult to incorporate into a real-time control scheme.  However, the data-driven method of dynamic mode decomposition advocated here, which is a regression to a best-fit linear model, is demonstrated to provide a viable path towards stable control of the coupled coil plasma system.

The diagnostics, circuits, and power supplies utilized on modern plasma physics experiments provide a wealth of data to which modern system identification techniques can be applied to learn the underlying dynamics of a particular interaction~\cite{humphreys2020}. In particular, linear reduced order models can be learned from time-series data, and then implemented in optimal linear feedback control and estimation. Techniques such as the {\em Dynamic Mode Decomposition} (DMD)~\cite{schmid_2010} allow for the discovery of such models, while also providing an interpretable (i.e. with explicit equations) representation of the dynamics. Optimal control and estimation has also been popular in plasma physics for some time, with examples in tokamak control~\cite{lazarus1990control, boyer2013first} being a particularly successful application of these techniques, as well as in plasma processing for semiconductor manufacturing \cite{rauf2002virtual, rashap1995control}. 

In recent years, there has been work showing the promise of interpretable data-driven system identification techniques in plasma physics. Nonlinear system-identified plasma models have been built for a number of complex plasma systems/dynamics, including: MHD simulations of helicity injection~\cite{kaptanoglu2021physics}, kinetic plasma simulations~\cite{alves2022data}, plasma closures~\cite{thevenin2022modeling}, plasma sources for semiconductor etching \cite{ko2023computational}, 2D electrostatic drift-wave turbulence~\cite{gahr2024learning}, turbulence bifurcations in fusion systems~\cite{dam2017sparse}, coupled plasma-neutral systems~\cite{Lore_2023}, and ablative pulsed plasma thrusters \cite{hossain2020reliable}. In addition, linear data-driven models have been built using the DMD for predicting the plasma dynamics in the HIT-SI line of devices~\cite{taylor2018dynamic,kaptanoglu2020characterizing}, kinetics prediction and acceleration~\cite{nayak2020dynamic,nayak2021detection,nayak2023accelerating}, and learning the dynamics of $E \times B$ plasmas~\cite{Faraji_2024_1,Faraji_2024_2}. Nonlinear and high-dimensional data-driven models with control, often using deep reinforcement learning, have also been successfully applied in recent years to learn control policies for tokamak shape design~\cite{degrave2022magnetic} and instability control~\cite{seo2024avoiding}.

\begin{figure}
  \centering
  \includegraphics[width=0.4\textwidth]{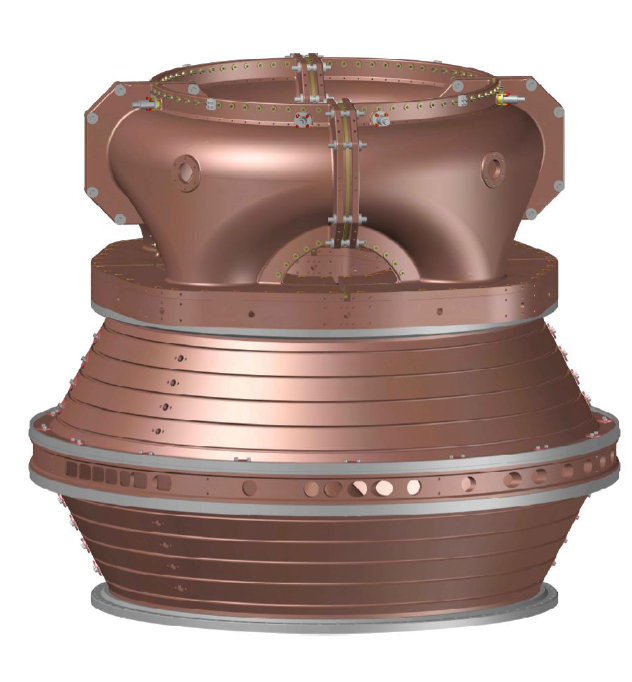}
  \includegraphics[width=0.9\linewidth]{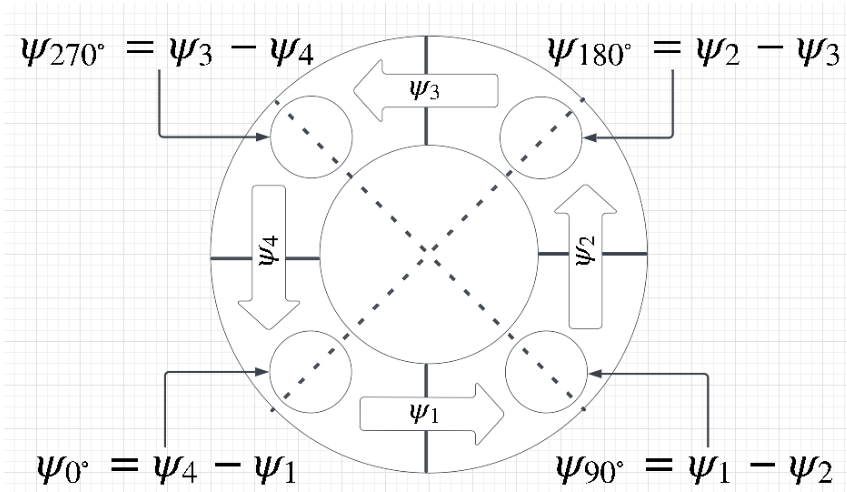}  
   \caption{Configuration of HIT-SIU injector manifold and associated fluxes.  The top figure shows a CAD drawing of HIT-SIU without coils, showing injector manifold (top) and confinement volume (bottom). A single injector consists of the path through the manifold between adjacent connections to the confinement volume.  A schematic of injector circuits on HIT-SIU as viewed looking down from above. $\Psi$ represents the magnetic flux through each injector, and the arrows represent the flow of magnetic flux around the manifold.}
  \label{fig:hitsiu-circuits}
\end{figure}

In this work we focus on linear modeling and control of the injector circuits of the {\em Helicity Injected Torus -- Steady Inductive Upgrade} (HIT-SIU), which are responsible for forming and sustaining a spheromak plasma inside the vacuum chamber. While these circuits exhibit a nonlinear coupling to the plasma, past work on the HIT-SI devices has shown that the nonlinear coil plasma system can be modeled as a linear system~\cite{kraske2014pid}. In addition, past work on tokamaks has shown that the nonlinear interaction between the plasma and the vertical control coils can be modeled and controlled as a linear system \cite{Kessel01051990}.  Further, we are interested in being able to easily modify an identified dynamics model without an expensive retraining phase, whether that be in terms of computational cost or accumulated training data. As such, the goal of this work is to implement the {\em bagging optimized dynamic mode decomposition} (BOP-DMD)~\cite{askham2018variable, sashidhar2022bagging} to discover linear models for the interaction between spheromak plasmas formed in HIT-SIU, and the helicity injectors used to form and sustain the plasma. BOP-DMD has a number of advantages over standard DMD, including (i) being optimally robust to noise, (ii) providing uncertainty quantification metrics, and (iii) allowing for constraining the linear model to be stable \cite{askham2018variable}. BOP-DMD models are then paired with optimal control and estimation, in the form of linear quadratic Gaussian control (LQG), with the objective of controlling the current profiles of the flux and voltage coils on the helicity injectors.  

The remainder of the paper is structured as follows: In Section II, we provide an overview of the HIT-SIU experiment, and steady inductive helicity injection (SIHI), in Section III we discuss the methodology used in this work, and in Section IV, we present results.

\section{\label{sec:level2}HIT-SIU}
\begin{figure}
    \centering
    \includegraphics[width=0.9\linewidth]{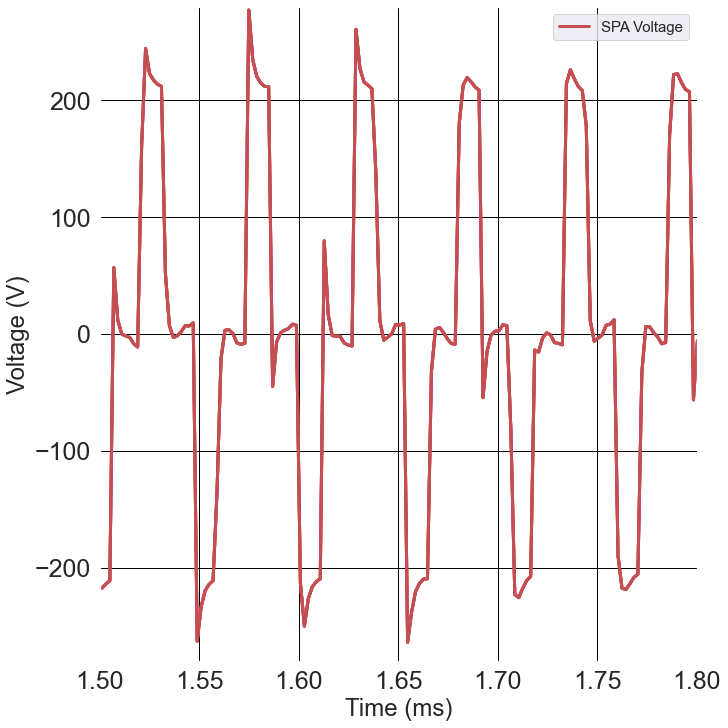}
    \caption{Typical SPA waveform for a shot on HIT-SIU. There is a clear maximum and minimum with a zero crossing time to ensure the SPA is not shorted. }
    \label{fig:spa_voltage}
\end{figure}

\begin{figure*}
    \centering
    \includegraphics[width=0.7\linewidth]{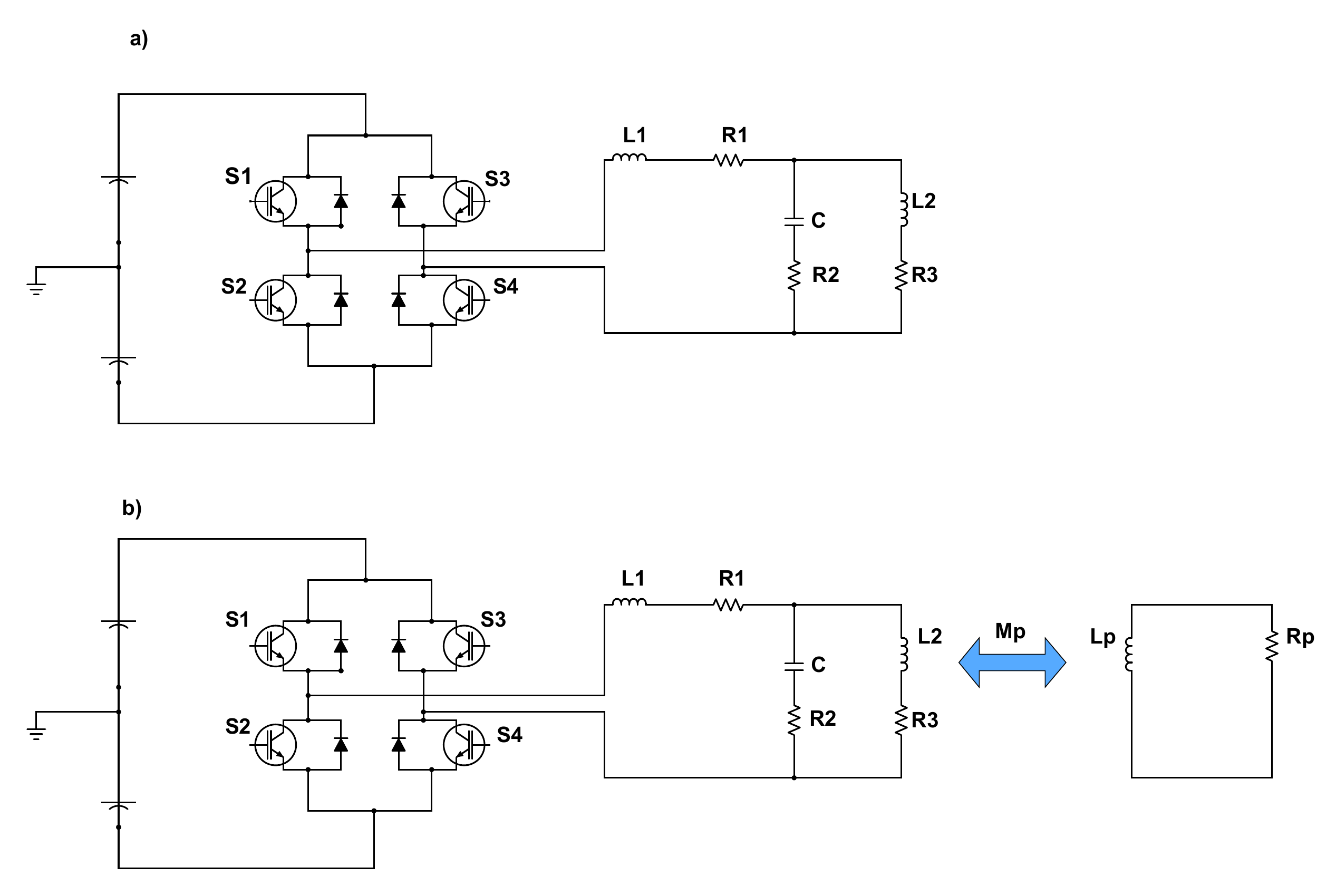}
    \caption{Circuit topology for a voltage or a flux circuit for a single injector. This diagram includes the SPA (S1-S4), the series inductor (L1), a parallel capacitance (C), and a parallel inductance (L2) which represents the voltage or the flux coil. a) is an example of a flux or voltage circuit in vacuum, and b) is flux or voltage circuit when a plasma is present. Lp denotes the inductance of the plasma, Rp its resistance, and Mp the mutual inductance between the circuit and the plasma. }
    \label{fig:circuit_topology}
\end{figure*}

The HIT-SIU experiment is studied using steady inductive helicity injection (SIHI)~\cite{jarboe1999steady,PhysRevLett.107.165005} to form and sustain spheromak plasmas. Energy and helicity are injected into the plasma using a set of four injectors, each comprised of a set of two coils, a so called voltage coil and a flux coil, that link a semi-toroidal channel that connects to the main plasma volume. In each injector the flux coil injects, locally toroidal, magnetic flux, while the voltage coil acts as an air core transformer generating a voltage parallel to the flux-coil-generated field. The magnetic fields of the voltage and flux coils are orthogonal to one another such that there is no mutual inductance between the two coils when no plasma is present. However, all four injectors are part of a common manifold structure (Fig.~\ref{fig:hitsiu-circuits}), so the four voltage coils, and separately all four flux coils, are inductively coupled. Each coil is part of a larger resonant circuit comprised of an additional inductor and capacitor, which is driven by a switching power amplifier (SPA). The SPAs drive a three-state square wave with a high, low, and zero crossing time. An example power supply waveform is shown in Fig.~\ref{fig:spa_voltage}. The injector circuit topology is shown on the left side of Fig.~\ref{fig:circuit_topology}. While the voltage and flux circuits have the same topology,  the values of each individual element differ between the two circuits. However, both circuits are tuned to the same resonant frequency.

\subsection{\label{sec:level2} Injector Circuit Control Methods:}
Fast, and accurate control methods are critical for creating, and sustaining, high-performance plasmas in present-day and future plasma physics experiments. The shape of the voltage and current waveforms for the HIT-SIU voltage and flux coils can determine the kinds of instabilities induced in the spheromak, as well as the overall helicity injected into the spheromak. All coils of the HIT-SIU experiment were controlled using a GPU-based proportional integral derivative (PID) controller~\cite{morgan2021high}. The PID algorithm parameterized the target waveforms to be driven as a sinusoid with an amplitude $V_0$ or $\Psi_0$, frequency $f_\text{inj}$, phase $\phi_A, \phi_B, \phi_C, \phi_D$, and offset, and used these control parameters to drive the desired current waveform through the coil. If the frequency and amplitude of the waveform driven on each injector is the same, the expression for the rate of helicity injection is given by
\begin{align}
   \dot{K} = 2V_{0}\Psi_{0}[&\sin^{2}(2\pi f_\text{inj}t + \phi_{A}) + \sin^{2}(2\pi f_\text{inj}t + \phi_{B})\\ + &\sin^{2}(2\pi f_\text{inj}t + \phi_{C}) + \sin^{2}(2\pi f_\text{inj}t + \phi_{D})]. \notag
\end{align}

\begin{figure}
    \centering
    \includegraphics[width=0.7\linewidth]{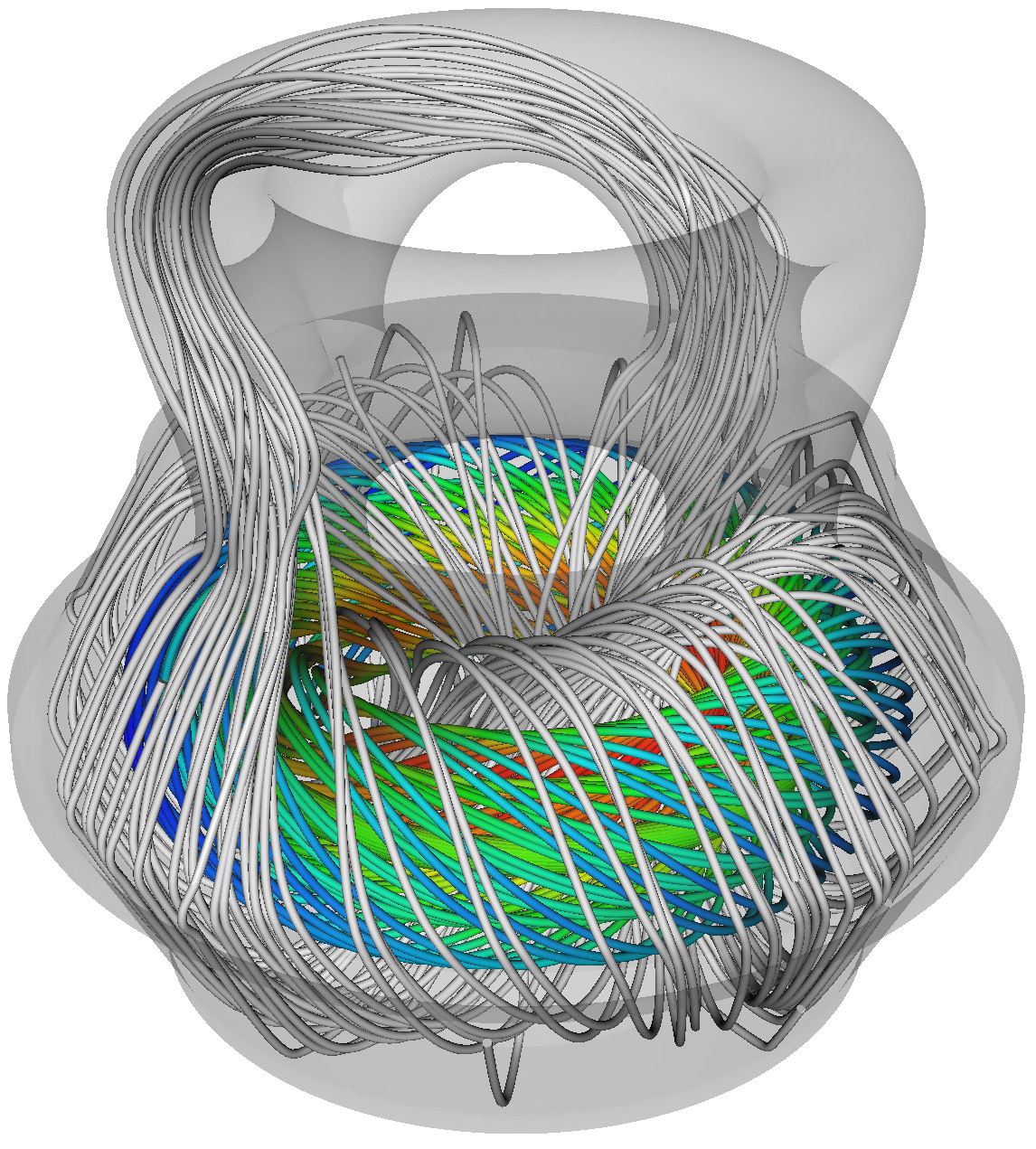}
    \caption{A representative equilibrium, with current gain of six, showing field lines linking (gray) and isolated (rainbow) from the injector volume. }
    \label{fig:hitsiu-Taylor}
\end{figure}

A typical shot begins with all four injectors in phase with one another, and then after breakdown occurs, the injectors are shifted out of phase with one another using the PID algorithm to achieve maximal helicity injection, or induce a particular structure in the plasma~\cite{morgan2022effect}. Typically a rotating structure with low toroidal mode number, as shown in Fig.~\ref{fig:hitsiu-Taylor} is targeted. However, by the end of the shot, the injectors can drift up to ten degrees out of phase with their desired trajectories, even with PID control. As the circuits drift out of phase, the rate of helicity injection decreases and the plasma's performance is adversely affected.  

While the PID algorithm implemented in ~\cite{morgan2021high} was a significant improvement upon the previous control scheme for the HIT-SI devices, further improvements can be made by shifting from model-free control to model-based control. 

Linear optimal control has been utilized across science and engineering to control complex systems~\cite{hespanha2018linear}. Optimal linear control involves pairing an optimal state estimator, which reconstructs the full state of the system from limited noisy measurements, with a feedback controller, which calculates the optimal input to the system to track a desired trajectory. This optimal estimator, when constructed from a solution to the algebraic Riccati equation, is known as a Kalman filter, and the feedback controller, when also found as the solution to the Riccati equation, is called the linear quadratic regulator (LQR). These methods will pair together for LQG control. Both the Kalman filter and LQR rely on an underlying linear dynamics model of the system so that the current state and optimal gain to apply can be computed.

\section{Analytic Model}
While there have been studies~\cite{PhysRevE.104.015206, 10.1063/1.4917476,10.1063/1.4919277,Hossack_2017} on the injector dynamics of the HIT-SI family of devices using MHD simulations and nonlinear reduced order models (ROMs), a linear model of the interaction between the plasma and the injector circuits for the purposes of control has not been developed. Before this study, the work of ~\cite{kraske2014pid}, which derived a linear state space model of the injector circuit dynamics on HIT-SI3 and implemented a Kalman filter, was the only significant effort.
To derive the linear model that will form the basis for the LQG control loop we seek to implement, we will begin with the circuit topology for the injector circuits as shown in Fig. ~\ref{fig:circuit_topology}. Treating the voltage across the capacitor, and the current through each coil as the states of our system, and then incorporating the mutual inductance between each flux or voltage coil, yields a 12-dimensional coupled system. This 12-dimensional system will be referred to as the vacuum circuit model. When a plasma is present one or more additional states need to be added to the model. In this work we choose to represent the plasma as a series LR circuit, which is coupled with a mutual inductance to each voltage or flux circuit, driven at or near its resonance. Therefore we elect to keep a fixed value for the inductance of the plasma. This yields a 13-dimensional coupled linear model. The full analytic vacuum model is available in Appendix~\ref{Appendix:analytic-vacuum-model}, and the analytic model for when a plasma is present is available in Appendix~\ref{Appendix:analytic-plasma-model}. 

There are of course many ways to represent the coupling between the injector circuits and the plasma as a linear system. For example, one may wish to add a capacitor to the circuit that represents the plasma. However, this would involve adding an additional voltage measurement to the state vector that represents the coupled injector-plasma system. By constructing a more complex circuit, a more accurate model of the coupled coil-plasma system may be derived, at the cost of additional model complexity and the inclusion of additional measurements. The methods proposed here are able to be adapted to any representation of the plasma as a linear circuit model.

The values used in the state space model for the inductance of the various circuit elements were obtained by applying a Hilbert transform to the voltage and current waveforms from each element during an experimental vacuum shot. However, there is a slight difference between the capacitance used in the circuit model, and the listed capacitance for the flux circuits in HIT-SIU. The capacitance for each flux circuit in the model range from 94.08 to 123.96 microfarads. These values were chosen by inspection to achieve better matching between the phases of the flux coil current waveforms of the model, and the observed flux coil current waveforms during vacuum shots. 

This model does make some important assumptions.  Specifically, we assume that when a plasma is present in the confinement volume, there is no inductive coupling between the plasma and the series inductor in each injector circuit. This assumption, which is well motivated as the series inductor is separated by a significant distance from the device, then allows us to conclude that the matrix that determines how actuation affects the dynamics of the injector circuits is constant between vacuum and plasma regimes. This can be observed from the structure of the $\mathbf{B}$ matrix shown in Appendix~\ref{Appendix:actuation-matrix}. By using standard state space techniques for circuit analysis, it is clear that the power supply input will only directly change the state of the series inductor. Thus, the structure of the circuit is such that we can isolate differences between the vacuum and plasma regimes to the flux or voltage coils.

\begin{figure*}
    \centering
    \includegraphics[width=1\linewidth]{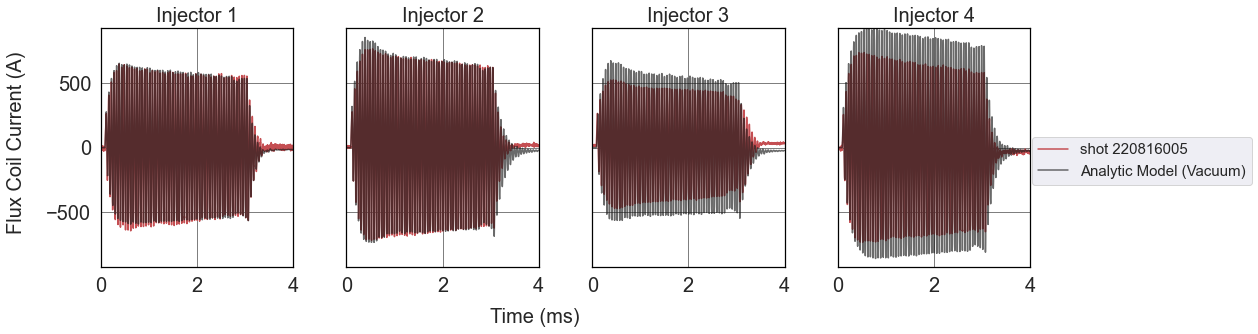}
    \caption{The analytic model of the flux circuits in vacuum is simulated using a power supply waveform for the fifth shot of August 16th 2022. The analytic model captures much of the circuit dynamics.}
    \label{fig:analytic_vs_vacuum_shot}   
\end{figure*}

\begin{figure*}
    \centering
    \includegraphics[width=1\linewidth]{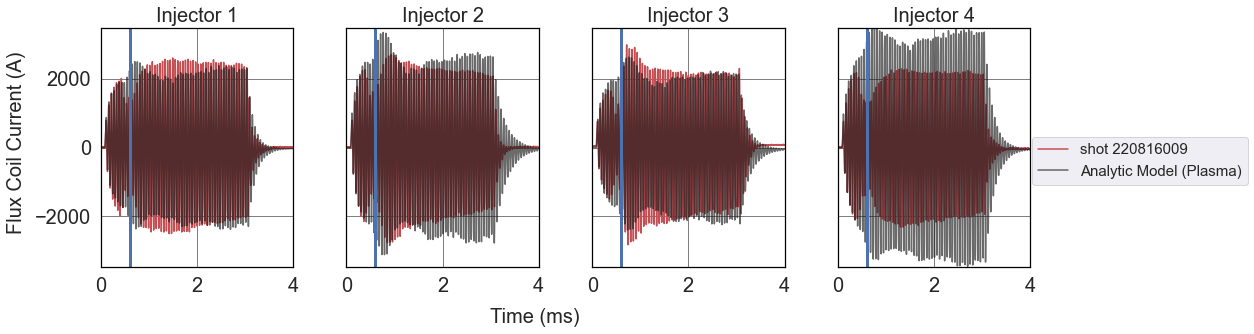}
    \caption{The analytic linear model for the flux circuits when a plasma is present is simulated using a power supply waveform from the ninth shot on August 16th, 2022. The analytic model represents the spheromak as an inductor that is mutually coupled to the other flux circuits. To the left of the vertical blue line ($\approx 1$ ms), there is no plasma present, and to the right, a plasma is present. }
    \label{fig:analytic_vs_plasma}
\end{figure*}

\subsection{\label{sec:level2}Matlab and Simulink Model:}
\begin{figure*}
    \centering
    \includegraphics[width=1\linewidth]{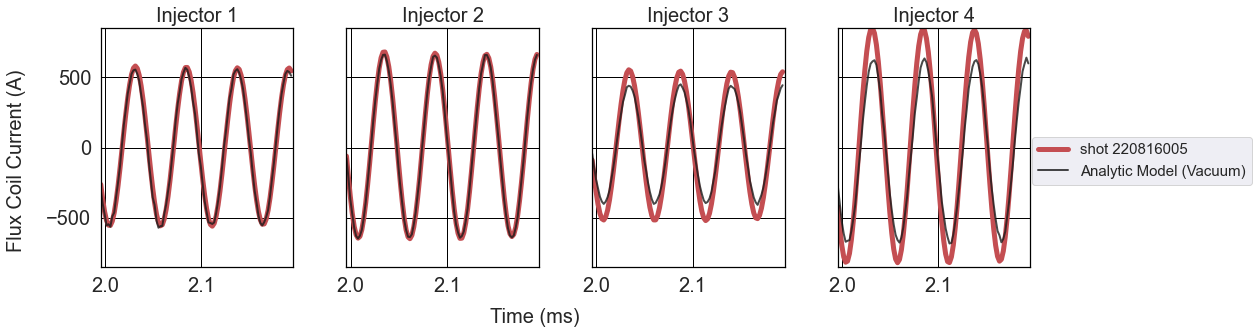}
    \caption{The analytic model of the flux circuits in vacuum matches the amplitude and phasing of current through the flux coils. }
    \label{fig:analytic_vs_vacuum_zoomed_in}
\end{figure*}

Developing control methods for hardware first requires a demonstration of performance in a digital environment. To that end, a model of the coupled flux circuits with full feedback control was developed in Matlab and Simulink. To verify the impedance of the various circuit elements, shot data was analyzed to capture the reactance and resistance of the elements, and to capture the strong and weak mutual inductance present between the injector circuits. Impedance is defined as \begin{equation}
\label{impedance-equation}
    \mathrm{Z} = R + iX.
\end{equation} The real part of Equation~\ref{impedance-equation} is the resistance of a circuit element, and the imaginary part of Equation~\ref{impedance-equation} is the reactance. For an inductor, the inductance is related to the reactances by $\mathrm{Im(Z)} = i\omega L$. For a capacitor, the capacitance is related to the reactance by $\mathrm{Im(Z)} = \frac{1}{i\omega C}$ ~\cite{alexander2017fundamentals}. The resistance that is found for each circuit element is represented as a resistor in series with the element as shown in Fig. ~\ref{fig:circuit_topology}. Since the vacuum model can be understood as a special case of the plasma model with no mutual inductance between the plasma and the flux coils, we use the same impedance for the flux circuit elements across models. However, the value chosen for the impedance of the plasma in the plasma model, the resistance in the plasma, and the coupling between the plasma and the flux coils, were chosen by inspection, and intuition about the behavior of the plasma. To evaluate the performance of the analytic vacuum and plasma models with respect to capturing the dynamics present during an experimental shot, an SPA waveform from a  shot is used as an input to the model, and the outputs of the model are evaluated against the flux circuit measurements from the experiment. The results of this test are shown in Fig.~\ref{fig:analytic_vs_vacuum_shot}, and Fig.~\ref{fig:analytic_vs_plasma} for the vacuum and plasma models respectively. Fig.~\ref{fig:analytic_vs_vacuum_zoomed_in} shows that the amplitude and phase of the injector circuits are well matched by the analytic model. While the analytic model is able to capture much of the behavior of the flux circuits in vacuum, the plasma model struggles to capture the behavior of the flux circuits when a plasma is present.

\begin{figure}
    \centering
    \includegraphics[width=1\linewidth]{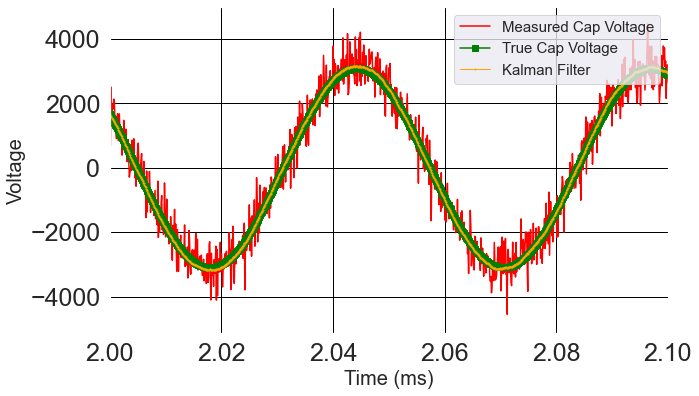}
    \caption{The Kalman filter has access to four noisy flux coil currents, and none of the other eight states, and is able to reconstruct the remaining states of the system based on the underlying model of the injector circuits in vacuum. Shown here is the reconstruction of the capacitor voltage from the first flux circuit.}
    \label{fig:kalman_filter_on_cap_data}
\end{figure}

With the derivation of a model that accurately captures the dynamics of interest in our system, the next step in the control scheme is to implement a Kalman filter. The Kalman filter uses a linear state space model of the injector circuit dynamics to estimate the state of the system from limited, and noisy measurements. We first demonstrate the performance of the Kalman filter when the underlying model is the analytic model for the injector circuits in vacuum, that was discussed previously in this section. Figure~\ref{fig:kalman_filter_on_cap_data} demonstrates the performance of the Kalman filter when reconstructing the voltage across the capacitor in the first flux circuit. While the filter only has access to four noisy measurements of the flux coil currents, it is still able to accurately predict the value of the capacitor voltage throughout the shot.

\begin{figure*}
    \centering
    \vspace*{-1in}
    \includegraphics[width=1\linewidth]{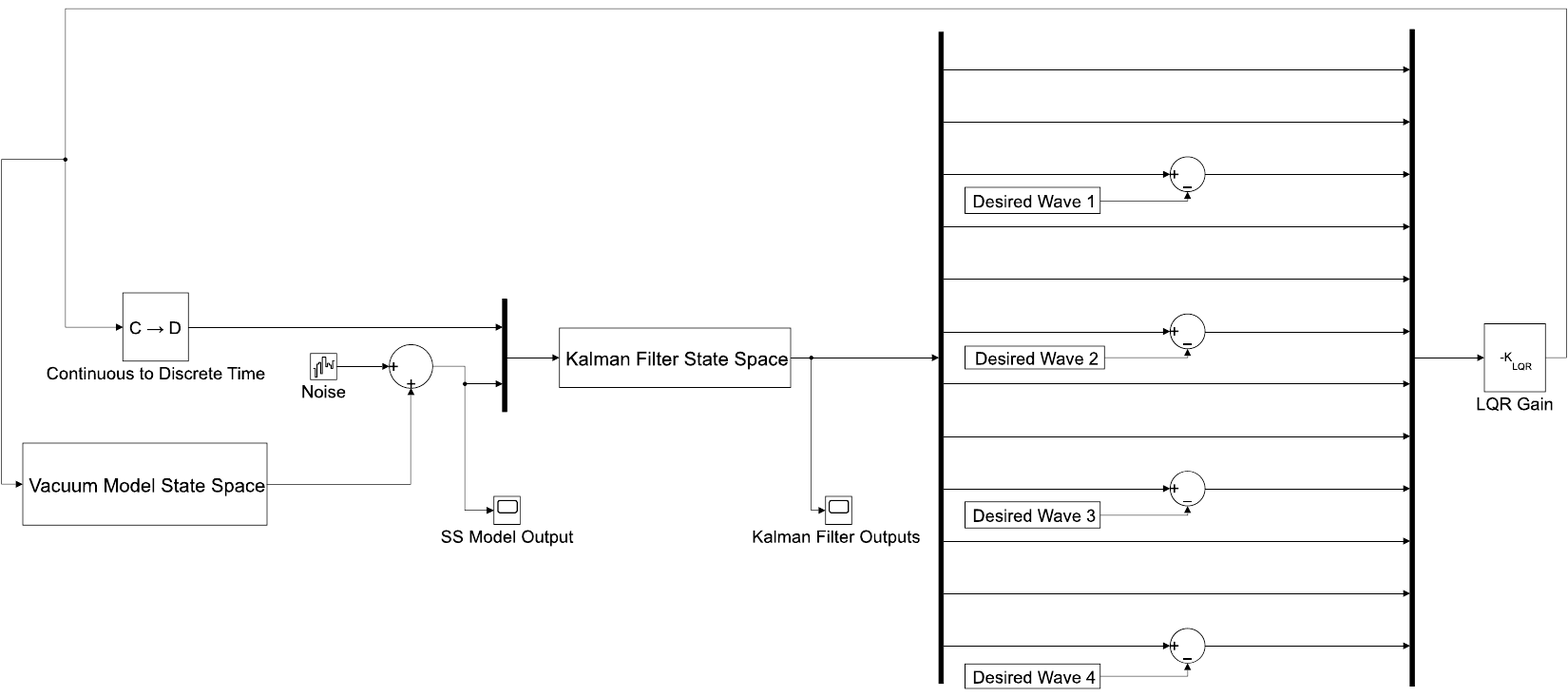}
    \vspace*{-1in}
    \caption{Full LQG feedback control loop in Simulink with the analytic model of the injector circuits in vacuum. The Kalman filter receives as input the states corresponding to the current through the flux coils, and from this measurement reconstructs the full state of the system. The desired flux coil current profiles are then subtracted off from the current state of flux coils, and fed into the LQR controller. }
    \label{fig:simulink model}
\end{figure*}

Finally, we contend with the controller: LQR minimizes a cost function $J$, that weighs the cost of control, with the value of producing an accurate reconstruction of the desired trajectory. These two costs are weighted by hyperparameters that can be tuned to prioritize accuracy of a specific state, and in this case was picked to be the current through the flux or voltage coil, as this trajectory is most closely linked to plasma performance. In this paradigm, a Kalman filter is used to predict from limited noisy measurements, the value of each state of the system at a given time. The LQR controller then makes a control decision to push the system to the desired state. The full control loop for controlling and estimating the injector circuits in vacuum is shown in Fig.~\ref{fig:simulink model}. 

\begin{figure}
    \centering
    \includegraphics[width=1\linewidth]{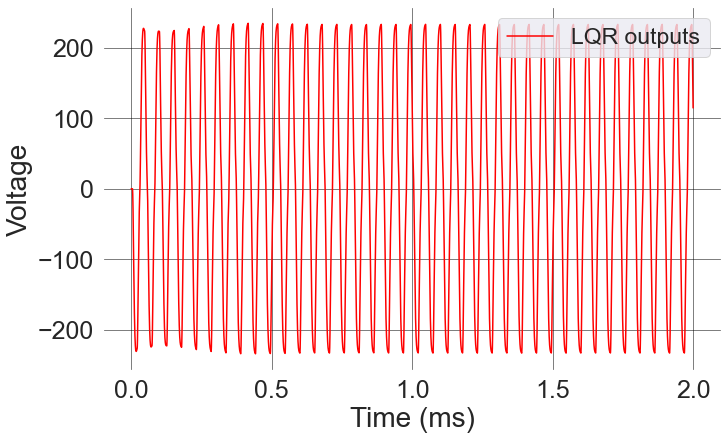}
    \caption{Voltage waveform determined by the LQR controller to best track the desired flux coil waveform for the injector circuits in vacuum.}
    \label{fig:LQR outputs}
\end{figure}

\begin{figure}
    \centering
    \includegraphics[width=1\linewidth]{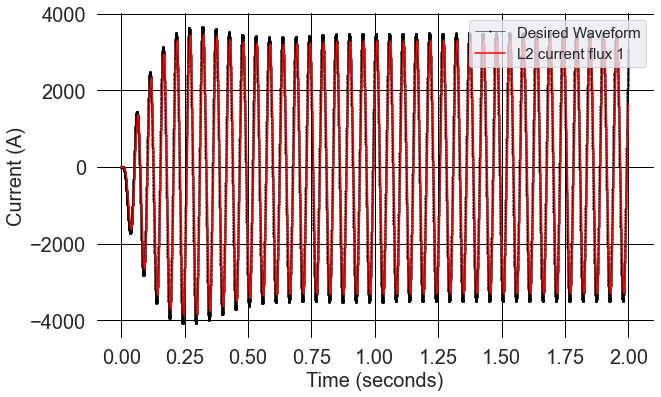}
    \caption{Output of the flux coil very closely matches the desired current waveform for the injector circuits in vacuum.}
    \label{fig:desired vs true output}
\end{figure}

A user must decide what the desired trajectory for the system is. HIT-SIU has three primary operating regimes that differ in the relative temporal phasing of the current in each flux coil. For this simulation, we choose the case where all circuits have the same temporal phasing, which is the simplest case of operation on HIT-SIU. The desired current waveform is arrived at by simulating the analytic model of the circuits with a standard power supply waveform seen on experimental shots. This desired waveform is then subtracted off from the flux circuit predictions that are the outputs of the Kalman filter. The LQR controller gain is then applied to the full state estimation with the desired trajectory subtracted off, allowing for trajectory tracking, instead of fixed point tracking. A power supply waveform generated by the LQR controller is shown in Fig.~\ref{fig:LQR outputs}. The action of the controller on the system produces a current profile through the flux circuit that closely matches the desired profile in Fig.~\ref{fig:desired vs true output}.

\section{Role of Data-Driven Models}
Data-driven model discovery and control are growing in popularity in plasma physics. As plasma physics devices increase in complexity, data-driven system identification will be a critical tool for understanding and controlling the complex, nonlinear phenomena they exhibit. While deep learning methods such as deep reinforcement learning have been applied with great success in plasma physics~\cite{degrave2022magnetic,seo2021feedforward}, these methods continue to struggle with interpretability, required training data, and uncertainty quantification~\cite{sashidhar2022bagging}. Learning interpretable governing equations allows for comparisons with models derived from first-principles. In addition, these methods can be easily paired with statistical techniques to obtain average coefficient values, as well as the variation in coefficients across trials supporting qualification in licensing and other contexts. These methods are also lightweight and require relatively low compute time, making them prime candidates for real-time control.
While there is some intuition about the nature of the evolution of the inductance on HIT-SIU, and previous iterations of the HIT-SI experiment, exact models for how the self and mutual inductance of the plasma evolve over time are not precisely known. However, there is a wealth of experimental data that can be used for the discovery of an improved linear model of the dynamics between the injector circuits and the plasma.

\subsection{The Dynamic Mode Decomposition}
The dynamic mode decomposition is one of the most common forms of linear system identification~\cite{schmid_2010}. The DMD is a purely data-driven algorithm for decomposing data into spatial modes with linear (exponential) time dynamics. These spatial modes oscillate at a fixed frequency, and grow or decay exponentially in time. Therefore DMD provides a linear dynamical system model for the spatiotemporal behavior of data. While the proper orthogonal decomposition/biorthogonal decomposition (POD/BOD) ~\cite{LIANG2002527,TDudokdeWitt_1995} provide the optimal basis for matrix approximation, these methods are unable to provide a general linear model for the time evolution of the data that can be extrapolated to new initial conditions. Exact DMD was originally developed to reconstruct flow fields in the fluids community, and we present the exact DMD formulation below ~\cite{H_Tu_2014}. 

DMD finds the best-fit linear operator that advances the dynamics forward in time,
\begin{equation}
\label{eq:one_step_linear_model}
    \mathbf{x}_{k+1} = \mathbf{Ax}_k.
\end{equation}
Where $\mathbf{x}_k$ is an $n$-dimensional vector, where $n$ is the dimension of the state space, and $A \in \mathbb{R}^{n \times n}$. To find this matrix $A$, we arrange measurements, often referred to as snapshots, into two matrices $\bm X$ and $\bm X'$. For $m$ measurements, $\bm X$,$\bm X'$ $\in \mathbb{R}^{n \times m}$  

\begin{align}    
    \mathbf{X} = \begin{bmatrix}
        |& |& & |\\
        x_0 & x_1 & ... & x_{m-1}\\
        |& |& & |
    \end{bmatrix}, \quad
    \mathbf{X'} = \begin{bmatrix}
        |& |& & |\\
        x_1 & x_2 & ... & x_{m}\\
        | & | & & |
    \end{bmatrix}.
\end{align}
These matrices are related through Eq.~\eqref{eq:one_step_linear_model},
\begin{equation} \label{eq:3}
    \mathbf{X'} = \mathbf{AX}.
\end{equation}
The best-fit linear operator that satisfies the relationship in Eq. (~\ref{eq:3}) is given as the solution to the matrix least squares problem
\begin{equation}
    \mathbf{A} = \argmin_{\mathbf{A}} ||\mathbf{X'} - \mathbf{AX}||_F = \mathbf{X'X}^{\dagger}.
\end{equation}
where $^{\dagger}$ represents the Moore-Penrose pseudoinverse, and $|| \cdot ||_F$ is the Frobenius norm. This can be rewritten in terms of the singular value decomposition (SVD) of $\mathbf{X}$
\begin{equation}
    \mathbf{X} = \mathbf{U}\boldsymbol{\Sigma}\mathbf{V}^*,
\end{equation}
where $\mathbf{U} \in \mathbb{C}^{n \times n}$, $\boldsymbol{\Sigma} \in \mathbb{R}^{n \times m}$, and $\mathbf{V} \in \mathbb{C}^{m \times m}$. Yielding
\begin{equation}
    \mathbf{A} = \mathbf{X'V} \boldsymbol{\Sigma}^{-1}\mathbf{U}^{*}.
\end{equation}
In the case where the full dimension of the system is large, it is computationally expensive to analyze $\mathbf{A}$ directly. Instead, $\mathbf{A}$ is projected onto the first $r \leq min(m,n)$ singular vectors of $\mathbf{U}$. This changes the dimensions of the matrices $\mathbf{U},\boldsymbol{\Sigma},\mathbf{V}$ such that $\mathbf{U}_r \in \mathbb{C}^{n \times r}$, $\boldsymbol{\Sigma}_r \in \mathbb{R}^{r \times r}$, and $\mathbf{V}_r \in \mathbb{C}^{m \times r}$, resulting in the approximate operator
\begin{equation}
    \mathbf{\Tilde{A}} = \mathbf{U}_{r}^{*}\mathbf{AU}_r = \mathbf{U}_{r}^{*}\mathbf{X'X}^{\dagger}\mathbf{U}_r = \mathbf{U}_{r}^{*}\mathbf{X'V}_{r}\boldsymbol{\Sigma}_{r}^{-1}.
\end{equation}

Having now solved for $\mathbf{\Tilde{A}} \in \mathbb{R}^{r \times r}$, the eigendecomposition of $\mathbf{\Tilde{A}}$ is computed to obtain the DMD eigenvalues and DMD eigenvectors
\begin{equation}
    \mathbf{\Tilde{A}W} = \boldsymbol{\Lambda}\mathbf{W},
\end{equation}
where the columns of $\mathbf{W}$ are the DMD eigenvectors, and $\boldsymbol{\Lambda} \in \mathbb{C}^{r \times r} $  is a diagonal matrix with entries $\lambda_j$ corresponding the $j$-th eigenvalue of $\mathbf{\Tilde{A}}$. The eigenvectors of the reduced $\mathbf{\tilde{A}}$ can be used to calculate the eigenvectors $\boldsymbol{\Phi}$ of the full $\mathbf{A}$ matrix, where
\begin{equation}
    \boldsymbol{\Phi} = \mathbf{X'W}\boldsymbol{\Sigma}^{-1}\mathbf{W}.
\end{equation}
Each eigen-pair corresponds to a distinct spatiotemporal mode, growing or damping at a rate $\mathrm{Re}(\lambda_j)$, and oscillating at a frequency $\mathrm{Imag}(\lambda_j)$. The states of the system can be approximated as a linear combination of the DMD modes
\begin{equation}
    \mathbf{x}_{k+1} \approx \sum_{i=1}^{r} b_i {\phi}_i\boldsymbol{\lambda}{_i}^{k} = \boldsymbol{\Phi} \mathbf{B} \boldsymbol{\Lambda}^k,
\end{equation}
where $B$ is the diagonal matrix containing the initial amplitudes of the DMD modes. $\mathbf{B}$ can be approximated as $\mathbf{B} =\boldsymbol{\Phi}^{\dagger}\mathbf{x_1}$. It is often easier to gain understanding of a DMD model by looking instead at the corresponding continuous time system. In this regime, the eigenvalues become $\omega_j = \log(\lambda_j)/\Delta t$ with frequency $v_j =\mathrm{Imag}(\omega_j)/2\pi$ and growth or damping of $\mathrm{z}_j = \mathrm{Re}(\omega_j)/2\pi$. In continuous time, the solution to the linear dynamical system becomes 
\begin{equation}
    \mathbf{x}(t) \approx \sum_{i=1}^r b_i{\bm{\phi}}_i \exp(\omega_i t) = \boldsymbol{\Phi} \mathbf{B}\exp(\boldsymbol{\Omega} t),
\end{equation}
where $\boldsymbol{\Omega}$ is a diagonal matrix, with the $j$-th diagonal entry corresponding to $\omega_j$. 

\subsection{The Optimized and Bagging Optimized DMD}

While the exact DMD has shown great promise for evenly sampled and noise-free data, the algorithm often fails to correctly identify modes in noisy data. The presence of sensor noise obscures the true relationship between consecutive snapshot pairs, leading to incorrect forecasts and often unstable models of the dynamics. There have been many efforts to improve upon exact DMD with respect to the effects of noise and actuation, such as forward-backward DMD~\cite{Dawson_2016}, Total-Least-Squares DMD~\cite{Hemati_2017}, Measure-preserving DMD~\cite{colbrook2022mpedmd}, Higher Order DMD~\cite{le2017higher} and DMD with control (DMDc)~\cite{proctor2016dynamic}. We adopt the Optimized DMD (OPT-DMD) from Askham and Kutz~\cite{askham2018variable} for performing DMD forecasting on noisy, experimental data as it offers the most robust framework to date for handling noise. Further, eigenvalue constraints for stable mode construction can be easily integrated into the framework.  DMD, OPT-DMD and BOP-DMD algorithms are all included in the pyDMD package~\cite{ichinaga2024pydmd}.

OPT-DMD reformulates the exact DMD as a nonlinear optimization problem and directly solves for the eigenvalues and eigenmodes of the DMD operator, denoted here as $\boldsymbol{\Omega} \in \mathbb{C}^{n \times n}$ and $\boldsymbol{\Phi_B} = \boldsymbol{\Phi B} \in \mathbb{C}^{n \times m}$,
\begin{equation}
    \min_{\boldsymbol{\Omega},\boldsymbol{\Phi}_B}||\mathbf{X} - \boldsymbol{\Phi}_B\exp(\boldsymbol{\Omega} t)\|_F.
\end{equation}
This procedure also allows for the user to constrain the learned model to be linearly stable, and  therefore avoid unstable forecasting. The physical injector circuits of HIT-SIU are linearly stable, so throughout the course of this work we enforce that the learned OPT/BOP DMD models are stable. Linear stability is enforced in the Matlab implementation of OPT-DMD that was developed for ~\cite{askham2018variable}. The optimizer fits an initial OPT-DMD model, and checks if any of the learned eigenvalues have a positive real part. If so, the positive real components of the eigenvalues are set to zero, while preserving the complex component. These new eigenvalues are then used as an initial guess for another call of OPT-DMD. 

OPT-DMD was further expanded upon by Sashidhar and Kutz~\cite{sashidhar2022bagging} to include statistical bagging techniques (BOP-DMD), adding further robustness to OPT-DMD, and allowing for uncertainty quantification of the learned modes, and eigenvalues. As outlined in ~\cite{sashidhar2022bagging}, a BOP-DMD model is computed by fitting many OPT-DMD models to randomly selected subsets of the training data, and then taking the mean over the eigenvalues, eigenvectors, and weights of each of these models to arrive at an "average" model of the dynamics. Means are taken component wise for vectors, and are taken separately for the real and complex components of the eigenvalues and eigenvectors. To ensure that eigenpairs that are trying to capture the same underlying physics are averaged with one another, the eigenpairs are sorted during each iteration of BOP-DMD. The authors of this work experimented with using medians as well, but found this did not increase the performance of BOP-DMD for this particular dataset.  

\section{Results}
We demonstrate the effectiveness of BOP-DMD on simulated data, vacuum shots from HIT-SIU, and finally plasma shots.  While BOP-DMD tends to handle experimental data better than any of the other DMD variants, the current implementation of the algorithm is unable to disambiguate the effects of control on the dynamics of the system of interest. For a typical HIT-SIU shot that lasts roughly four milliseconds, there is one millisecond before the end of the shot where the power supplies are turned off and a portion of the dynamics are available for analysis without the effects of control. If one were to apply BOP-DMD to an actuated portion of shot, the DMD operator would encapsulate the effect of actuation on the dynamics, rather than capturing the unforced dynamics.

Due to the nature of the power supply waveform, the response of the circuits for this last millisecond resemble their impulse response, providing a perfect testing ground for training a BOP-DMD model. By training on this quasi impulse response, we are able to obtain a model for the dynamics of the circuits in vacuum that is able to be generalized to other shots taken on that day.  After training a BOP-DMD model on the last millisecond of the shot, we evaluate the model by inputting the full power-supply waveform for the given shot, and compare between the BOP-DMD model and the actual response of the circuits to this waveform. We also evaluate the performance of a model trained on one shot on other shots taken on that day.

After extensive testing, we have found that including nine DMD modes for a plasma model, and five DMD modes for a vacuum model, provide the best models for reconstruction of the training shot, and the prediction of future shots.

\subsection{BOP and OPT DMD on Simulated data}
\begin{figure}
    \centering
   
    \includegraphics[width=1\linewidth]{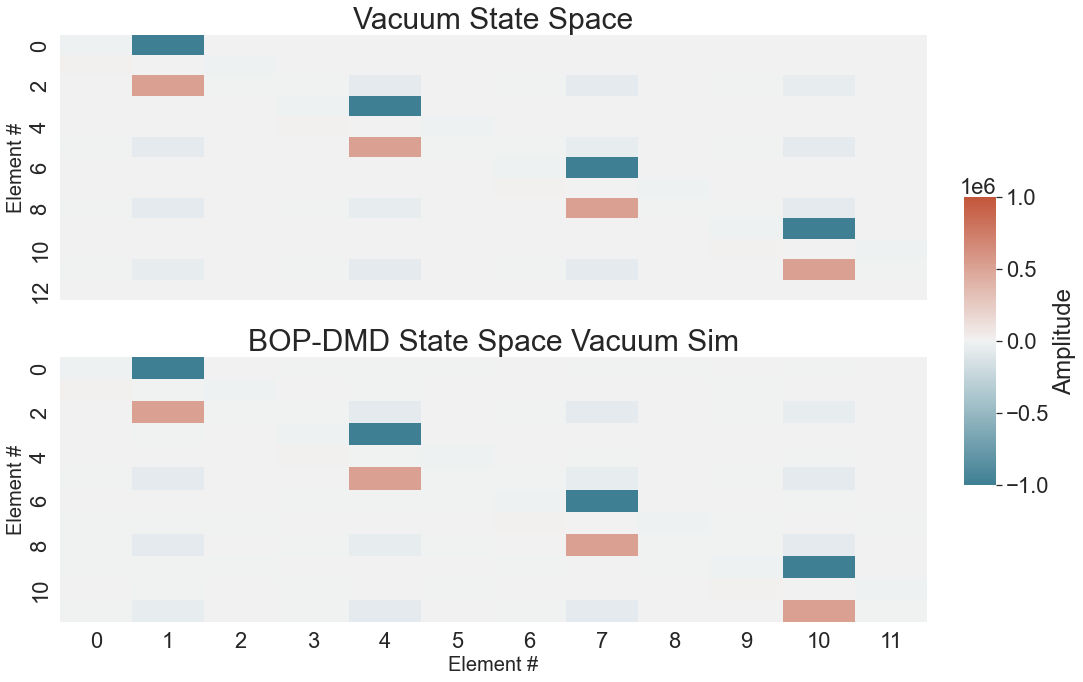}
    \caption{By introducing statistical bagging, BOP-DMD correctly learns the dynamics matrix. This was trained on the last quarter of a simulated vacuum shot when the power supplies had turned off. 10 trials, or bags, were used for this model.}
    \label{fig:bopdmd vs vacuum ss from simulated data heatmap}
\end{figure}

\begin{figure}
    \centering
    \includegraphics[width=1\linewidth]{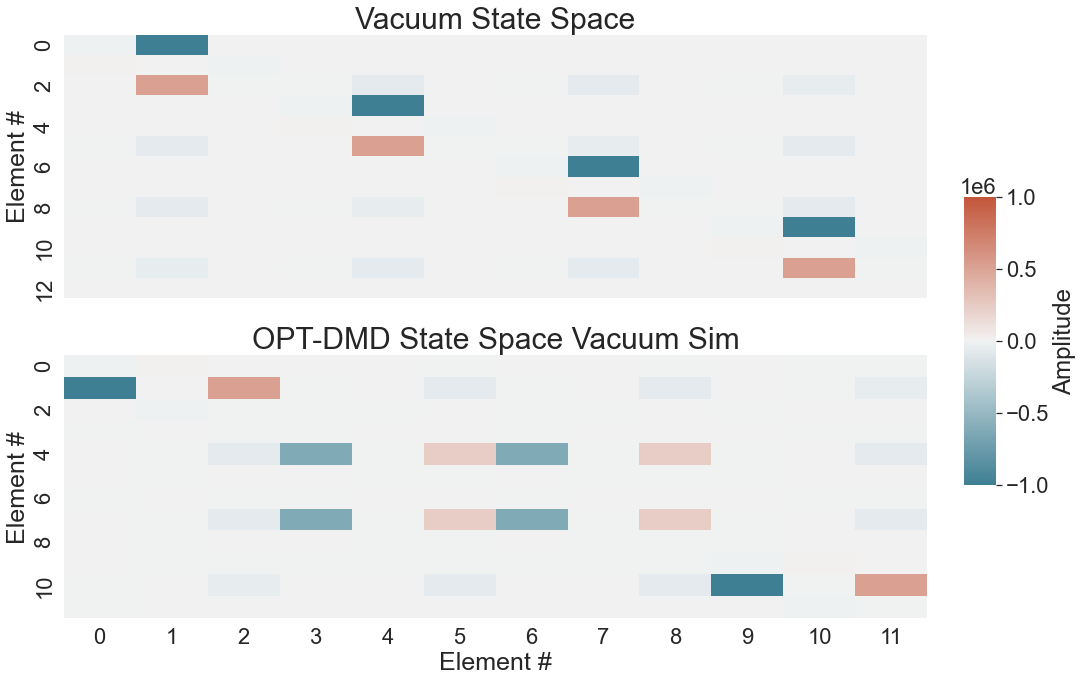}
    \caption{OPT-DMD is trained on the last quarter of a simulated vacuum shot once the power supplies have turned off. There are spurious entries to the DMD operator computed by the OPT-DMD that are able to be rectified with the addition of statistical bagging.} 
    \label{fig:optdmd_heatmap}
\end{figure}

To illustrate the advantages of BOP-DMD over OPT-DMD, we train BOP-DMD and OPT-DMD models on noiseless  circuit data generated by simulating the analytic model derived in Appendix~\ref{Appendix:analytic-vacuum-model}. Training data was generated by simulating the linear model of the vacuum circuits that was presented in Section II, with a voltage waveform designed to match the experimental waveforms generated by the SPAs on HIT-SIU. The circuits were simulated for four milliseconds with the last millisecond of the voltage waveform being set to zero to once again match experimental conditions. While both OPT-DMD and BOP-DMD were able to correctly reproduce training data, and both performed well on test data, one can see the effects of bagging by examining the entries of the linear operator learned by the respective methods. As shown in Figures~\ref{fig:bopdmd vs vacuum ss from simulated data heatmap} and~\ref{fig:optdmd_heatmap}, by introducing statistical bagging, BOP-DMD is able to correctly identify the true structure of the operator that generated the training data, while OPT-DMD learns spurious entries to this matrix.

\subsection{BOP-DMD on Vacuum Shots}
\begin{figure*}
    \centering
    \includegraphics[width=1\linewidth]{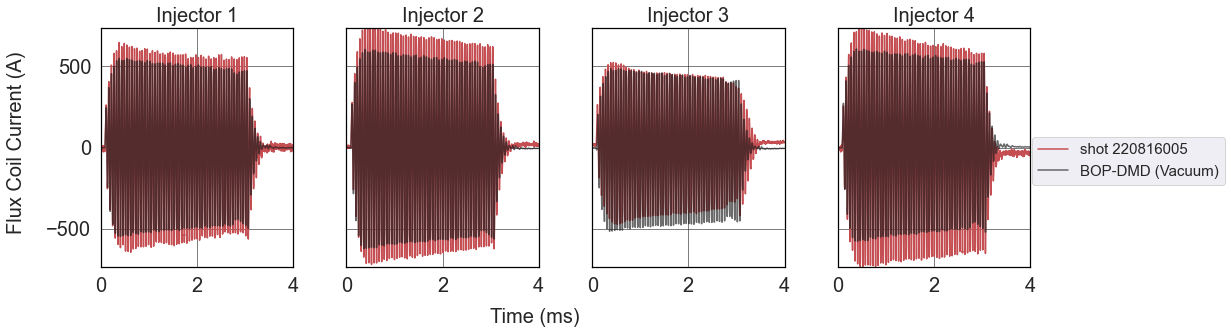}
    \caption{BOP-DMD trained on the fifth shot of August 16, 2022 in vacuum. This model was trained on the last millisecond of the shot once the power supplies had turned off, and then evaluated by simulating the model with the full power supply waveform comparing to the original shot. The model shown used five DMD modes (much less than the full 12 modes of the analytic model), and used 20 bags.}
    \label{fig:BOP_vacuum_shot220816005}
\end{figure*}

\begin{figure*}
    \centering
    \includegraphics[width=1\linewidth]{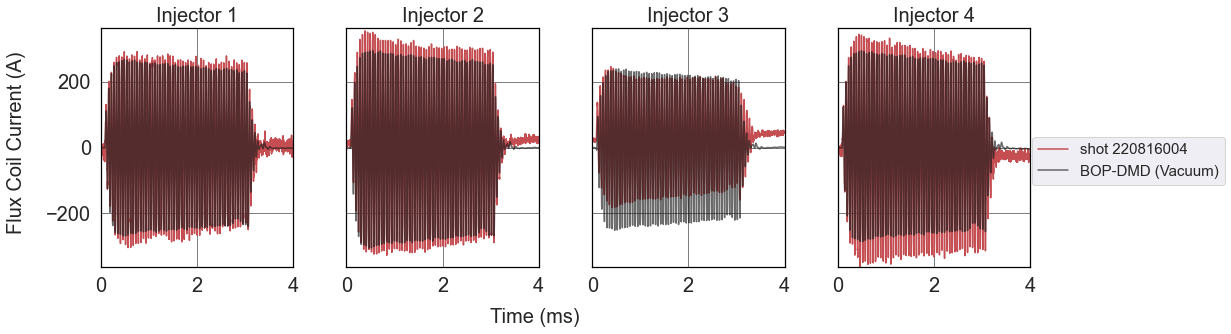}
    \caption{This is the same model as used in Fig. ~\ref{fig:BOP_vacuum_shot220816005}, but is now simulated using a power supply waveform from the previous shot in vacuum (the fourth shot of August 16th, 2022). This model generalizes from shot to shot from the same day. }
    \label{fig:bop model on new vacuum shot}
\end{figure*}

When beginning to examine the methodology outlined above, we initially tested the capabilities of BOP-DMD on data that was known to be generated by a linear dynamical system: the flux circuits in vacuum. Coupled LC circuits are a classical linear dynamical system, and thus present a test-bed for the performance of BOP-DMD under real-world experimental conditions. The results of BOP-DMD trained on this kind of data can be seen in Figs.~\ref{fig:BOP_vacuum_shot220816005} and~\ref{fig:bop model on new vacuum shot}. In Fig.~\ref{fig:BOP_vacuum_shot220816005}, we show the performance of BOP-DMD trained on the last millisecond of a vacuum shot as the power supplies turn off, and then evaluated on the entire shot by feeding in full power supply waveform. The model trained in Fig.~\ref{fig:BOP_vacuum_shot220816005} is then evaluated on the previous vacuum shot from that day Fig.~\ref{fig:bop model on new vacuum shot}. These results indicate that BOP-DMD is able to learn the underlying dynamics of the vacuum circuits, while not being subject to over fitting. Further, BOP-DMD is able to learn the correct structure of an underlying operator, and reduce the impact of spurious entries through statistical bagging.

\subsection{BOP-DMD With a Plasma Present}
\begin{figure*}
    \centering
    \includegraphics[width=1\linewidth]{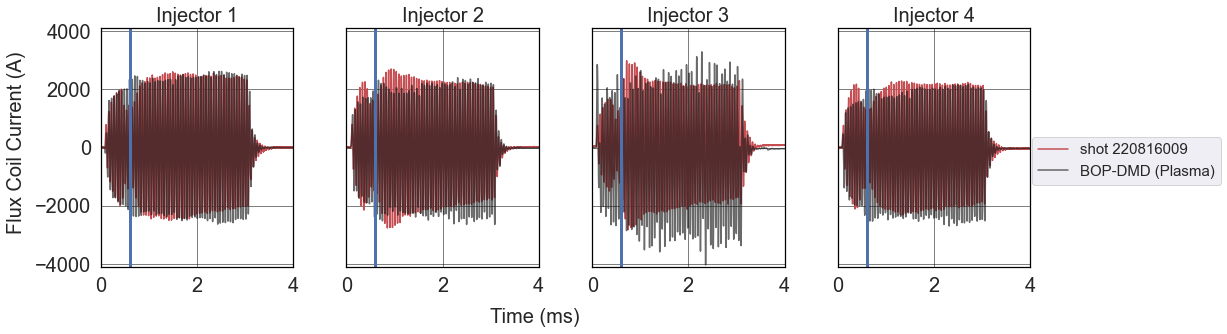}
    \caption{BOP-DMD trained on the last millisecond of the 9th shot of August 16th, 2022. This is a shot when a plasma is present. Even though BOP-DMD is trained on the final transient at the end of the shot when the spheromak is decaying, the model is able to capture much of the flux circuit dynamics in the beginning of the shot. This model utilizes nine DMD modes. To the left of the vertical blue line ($\approx .6$ ms), there is no plasma present, and to the right, a plasma is present.}
    \label{fig:BOP on plasma shot}
\end{figure*}

\begin{figure*}
    \centering
    \includegraphics[width=1\linewidth]{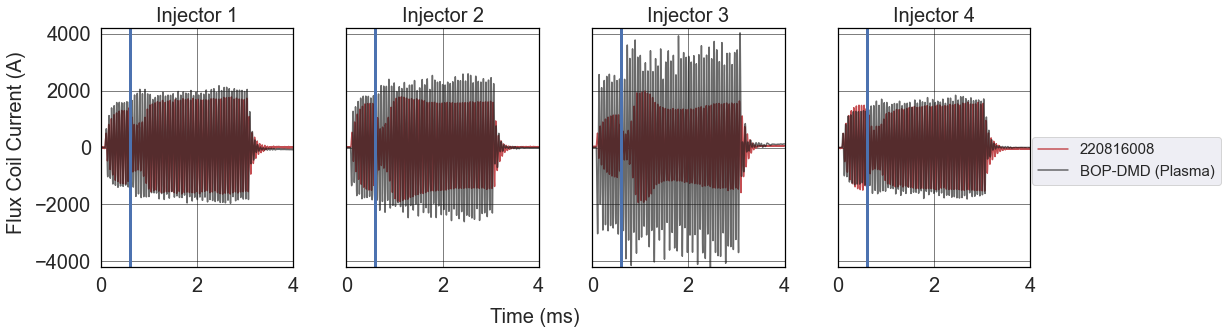}
    \caption{BOP-DMD model trained on last millisecond of shot 220816009 simulated with a new power supply waveform from the previous shot and evaluated against the ground truth. Even though the BOP-DMD model had no training data from this shot, the model still captures the flux circuit dynamics. To the left of the vertical blue line ($\approx .6$ ms), there is no plasma present, and to the right, a plasma is present.}
    \label{fig:BOP on new plasma shot}
\end{figure*}

\begin{figure}
    \centering
    \includegraphics[width=1\linewidth]{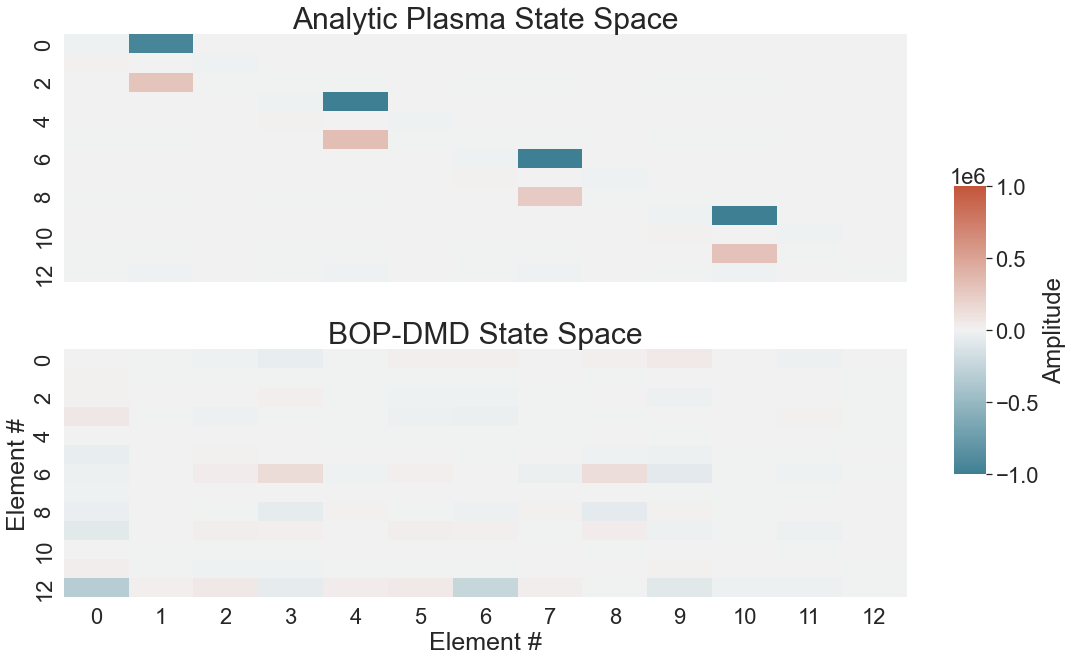}
    \caption{Comparison of the structure of BOP-DMD $\mathbf{A}_p$ matrix and analytic plasma model.}
    \label{fig:comparison of plasma ss and bop dmd from plasma shot heatmap}
\end{figure}

\begin{figure}
    \centering
    \includegraphics[width=1\linewidth]{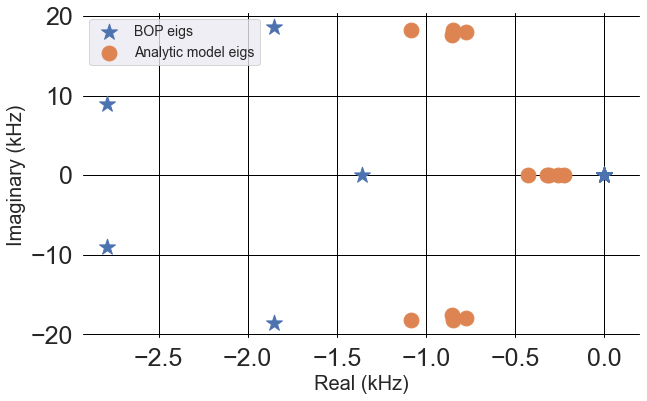}
    \caption{Eigenvalues of the analytic plasma model compared with the eigenvalues of the BOP-DMD model trained on shot 220816009. }
    \label{fig: eigenvalue comparison}
\end{figure}

\begin{figure}
    \centering
    \includegraphics[width=1\linewidth]{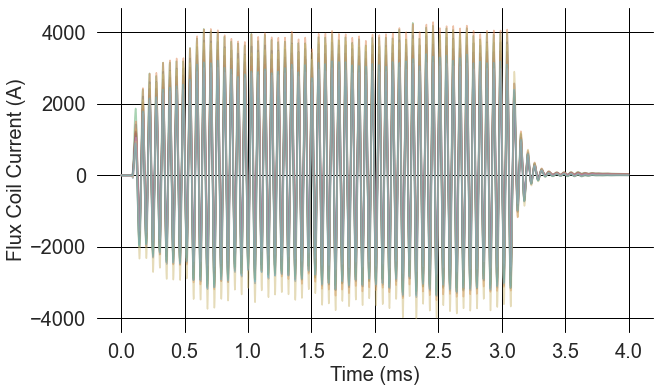}
    \caption{Prediction of the flux coil current through the first flux circuits from the first, second and fourth models from the BOP-DMD model that was trained on shot 220916009. }
    \label{fig:individual bop trajectories}
\end{figure}

We now evaluate our methodology on shots where a plasma is present. By augmenting the states of our system to also include the toroidal plasma current, we train a BOP-DMD model on the last millisecond of a plasma shot. During this final millisecond of the shot, the toroidal plasma current decays from it's steady state behavior as the plasma in the confinement volume begins to dissipate. Again, due to the limitations of BOP-DMD, this methodology is unable to access the steady state interaction between the flux circuits and the spheromak. However, this method still provides reasonable results; Fig.~\ref{fig:BOP on plasma shot} shows an example of a BOP-DMD model fit on the last millisecond of a plasma shot and then evaluated on the rest of the shot. Further, this model is able to be applied to other plasma shots taken on the same day Fig.~\ref{fig:BOP on new plasma shot}. However, the error between the predicted current and the measured current in the third flux coil increases between the training shot, and the test shot. Because of the small differences in the relative phasing of the power supplies between the two shots, there is a different mutual inductance between the circuits and the plasma. As previously mentioned, one of the primary limitations of the BOP-DMD method is the inability to disambiguate the effects of actuation on the dynamics of a system. This leads to a fundamental gap in the training data available to the method for capturing the coupling between the injectors circuits and the plasma. Future work should focus on not only incorporating actuation into BOP-DMD, but also stringing together multiple shots for a wider array of training data. We do notice however that the operator learned by BOP-DMD does not resemble the same sparse structure that is present in the analytic plasma model Fig.~\ref{fig:comparison of plasma ss and bop dmd from plasma shot heatmap}. Despite losing the sparse structure of the dynamics matrix, the modes of the BOP-DMD model oscillate at similar frequencies to the analytic model, despite decaying faster in general. From Fig. ~\ref{fig: eigenvalue comparison}, one can see that both the analytic model, and the BOP-DMD model identify dominant modes at roughly 20 kHz, which matches the resonant frequency of the injector circuits. It is also useful to examine the different predictions provided by each individual OPT-DMD model in the 20 iterations of BOP-DMD ~\ref{fig:individual bop trajectories}.

\begin{figure}
    \centering
    \includegraphics[width=1\linewidth]{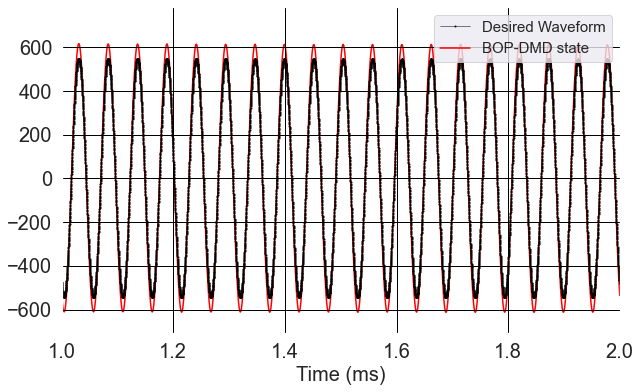}
    \caption{The LQR controller is able to push the BOP-DMD model to accurately track the desired trajectory. }
    \label{bop lqg}
\end{figure}

\subsection{Implementation of BOP-DMD in a Control Loop}
Now that we have evaluated the ability of BOP-DMD to learn linear models of the flux circuits both in the vacuum case and the plasma case, we will evaluate the effectiveness of this model as a plant in an LQG feedback loop. Because this model is of reduced rank , we begin with taking an SVD of the BOP-DMD matrix, and projecting onto the first five columns of $\mathbf{U}$,
\begin{align}
     \mathbf{A}_\text{DMD} &= \mathbf{U}\boldsymbol{\Sigma}\mathbf{V}^*, \\ 
    \mathbf{\tilde{A}}_\text{DMD} &= \mathbf{U}_{r}^*\mathbf{A}_\text{DMD}\mathbf{U}_r.
\end{align}
We then introduce a reduced state vector $\mathbf{z}$ such that \begin{equation}
    \mathbf{x} = \mathbf{U_{r}z}.
\end{equation}
This also changes the actuation matrix, and observation matrix as follows:
\begin{align}
    \mathbf{\tilde{B}} &= \mathbf{U}_{r}^*\mathbf{B}\\
    \mathbf{\tilde{C}} &= \mathbf{C}\mathbf{U}_{r}
\end{align}
Using this substitution, we arrive at reduced dynamics matrices $\mathbf{\tilde{B}}$, $\mathbf{\tilde{C}}$, as well as reduced noise covariance matrices, and cost matrices for the Kalman filter and LQR controller respectively. These new matrices were then used to design an LQG loop in Matlab and Simulink for the BOP-DMD model. As shown in Fig.~\ref{bop lqg}, the LQG loop is able to accurately track a desired trajectory. 

\section{Conclusion}
In this work, we have demonstrated the advantages of linear control for the flux and voltage circuits on the HIT-SIU device, and the role BOP-DMD plays in discovering linear models of the complex interaction between the flux and voltage circuits and the spheromak. While BOP-DMD has limitations in terms of disambiguating the effects of actuation on data, there are still many advantages to using BOP-DMD over first-principles linear models. BOP-DMD does not require knowledge of often hard-to-measure system parameters (in this case the inductance, mutual inductance, and capacitance of the various circuit components), and BOP-DMD is able to be tuned to shot-to-shot experimental conditions with ease. BOP-DMD's ability to provide interpretable, linear models of often nonlinear circuit behavior with uncertainty quantification make it a powerful tool for use in linear optimal control and estimation. 

We have also demonstrated in this work the advantages of linear optimal control to model-agnostic control schemes such as PID control, on the HIT-SIU device. By pairing linear optimal control, with fully data-driven system identification, we have demonstrated that a model-based control scheme can accurately model the flux circuit dynamics, and control the behavior of the model in a simulation environment. Despite the advancements put forth in this work, the authors acknowledge room for improvement in this methodology, specifically with regards to the need for DMD variants that are able to robustly remove the affect of actuation on a dynamical system in the presence of noise. This work highlights the need for new approaches such as a combination of BOP-DMD and DMDc, that leverages reformulating the classical DMD problem as a best fit of complex exponentials, while at the same time being able to remove the affect of a known control signal from the data. 

\section{Acknowledgments}
The authors would like to thank Nicholas Zolman, Michelle Hickner, and other members of the Data-Driven Dynamics and Control Laboratory at the University of Washington for valuable discussions. 

The information, data, or work presented herein was
funded in part by the National Science Foundation under Grant No. PHY-2329765 and the Advanced Research Projects Agency-Energy (ARPA-E), U.S. Department of Energy, under Award Number DE-AR0001266 and by CTFusion, Inc., the primary recipient of ARPA-E award number DE-AR0001098. 

S.L.B and J.N.K acknowledge support from the US National Science Foundation (NSF) AI Institute for Dynamical Systems (dynamicsai.org), grant 2112085.

\appendix
\section{Full Linear Model of Flux/Voltage Circuits}
\label{Appendix:analytic-vacuum-model}
The full vacuum and plasma circuit models were derived by assuming the voltage and flux circuits follow the circuit topology shown in Fig.~\ref{fig:circuit_topology}. For the plasma model, this circuit is modified to include the plasma as an inductor that is mutually coupled to the voltage or flux coils. The equations were solved numerically using Matlab's backslash command, and symbolically using Wolfram Mathematica's Solve command. The linear model for the circuits when a plasma is present was only solved using backslash, but it was verified that these equations reduce to the vacuum model as the inductance between the circuits and the plasma tended to zero. The linear equation matrix in vacuum is as follows:
\medmuskip=2mu
\thickmuskip=2mu
\thinmuskip=2mu
\begin{align}\notag
    A_v &= \begin{bmatrix}
        \frac{-(R_1 + R_2)}{L_1} \!\!\!\!\!\!\!\!\!\!\!\! & \frac{-1}{L_1}  \!\!\!\!\!\!\!\!\!\!\!\!& \frac{-R_2}{L_1}  \!\!\!\!\!\!\!\!\!\!\!\!& 0 \!\!\!\!\!\!\!\!\!\!\!\! & \dots \!\!\!\!\!\!\!\!\!\!\!\! & 0 \\
        \frac{1}{C}\!\!\!\!\!\!\!\!\!\!\!\! & 0\!\!\!\!\!\!\!\!\!\!\!\! & \frac{-1}{C} \!\!\!\!\!\!\!\!\!\!\!\! & 0\!\!\!\!\!\!\!\!\!\!\!\! & \dots\!\!\!\!\!\!\!\!\!\!\!\! & 0 \\
        -K x_{3,1}\!\!\!\!\!\!\!\!\!\!\!\! & -K x_{3,2}\!\!\!\!\!\!\!\!\!\!\!\! & -K x_{3,3}\!\!\!\!\!\!\!\!\!\!\!\! & -K x_{3,4}\!\!\!\!\!\!\!\!\!\!\!\! & \dots\!\!\!\!\!\!\!\!\!\!\!\! & -K x_{3,12}\\
        \vdots\!\!\!\!\!\!\!\!\!\!\!\! & \vdots\!\!\!\!\!\!\!\!\!\!\!\! & \vdots\!\!\!\!\!\!\!\!\!\!\!\! & \vdots\!\!\!\!\!\!\!\!\!\!\!\! & \vdots\!\!\!\!\!\!\!\!\!\!\!\! & \vdots\\
        0\!\!\!\!\!\!\!\!\!\!\!\! & 0\!\!\!\!\!\!\!\!\!\!\!\! & \dots\!\!\!\!\!\!\!\!\!\!\!\! & \frac{-(R_1 + R_2)}{L_1} \!\!\!\!\!\!\!\!\!\!\!\! & \frac{-1}{L_1}\!\!\!\!\!\!\!\!\!\!\!\! & \frac{-R_2}{L_1}\\
        0\!\!\!\!\!\!\!\!\!\!\!\! & 0\!\!\!\!\!\!\!\!\!\!\!\! & \dots\!\!\!\!\!\!\!\!\!\!\!\! & \frac{1}{C}\!\!\!\!\!\!\!\!\!\!\!\! & 0\!\!\!\!\!\!\!\!\!\!\!\! & \frac{-1}{C}\\
    -K x_{12,1} & -K x_{12,2} & -K x_{12,3} & -K x_{12,4} & \dots & -K x_{12,12}\\
    \end{bmatrix},\\
    K &= \frac{1}{(L_2-M_w) ( L_2^2 - 4 M^2 + 2 L_2 M_w+ M_w^2)}.
\end{align}
Here $K$ is a pre-factor, $L_1$ is the inductance of the series inductor and $R_1$ is the corresponding resistance, $C$ is the capacitance of the parallel capacitor and $R_2$ is the corresponding resistance, $L_2$ is the inductance of the flux or voltage coil and $R_3$ is the corresponding resistance, $M$ is the mutual inductance between a circuit and a nearest neighbor, and $M_w$ is the inductance between a circuit and its farthest neighbor. Due to the symmetry in the voltage and flux circuits, the structure of the first two lines repeats, only shifting over by three positions for each new circuit. The third line repeats with the pre-factor $K$ being negative in rows three and 12, and positive in rows 6 and 9. Each term in rows 3, 6, 9, and 12 are listed below.

% \begin{figure*}
\medmuskip=-2mu
\thinmuskip=-2mu
\thickmuskip=-2mu
\allowdisplaybreaks
\footnotesize{
\begin{align}
    x_{3,1} &=  -L_2^2 R_2+2 M^2R_2-L_2 M_w R_2 ,\\\notag
    x_{3,2} &= -L_2^2+ 2 M^2-L_2 M_w,\\\notag
    x_{3,3} &= L_2^2 R_2-2 M^2 R_2+L_2 M_w R_2+L_2^2 R_3-2 M^2 R_3+L_2 M_w R_3,\\\notag
    x_{3,4} &= L_2 M R_2 - M M_w R_2,\\\notag
    x_{3,5} &= L_2 M-M M_w,\\\notag
    x_{3,6} &= -L_2 M R_2+M M_w R_2-L_2 M R_3+M M_w R_3,\\\notag
    x_{3,7} &= L_2 M R_2-M M_w R_2,\\\notag
    x_{3,8} &= L_2 M-M M_w,\\\notag
    x_{3,9} &= -L_2 M R_2+M M_w R_2-L_2 M R_3+M M_w R_3,\\\notag
    x_{3,10} &= -2 M^2 R_2+L_2 M_w R_2+M_w^2 R_2,\\\notag
    x_{3,11} &= -2 M^2+L_2 M_w+M_w^2,\\\notag
    x_{3,12} &= 2 M^2 R_2-L_2 M_w R_2-M_w^2 R_2+2 R_3M^2-L_2 M_w R_3-R_3 M_w^2\\    \notag
% \end{align}
% \begin{align}
% \notag
x_{6,1} &= -L_2 M R_2+M M_w R_2,\\\notag
x_{6,2} &= -L_2 M+M M_w,\\\notag
x_{6,3} &= L_2 M R_2-M M_w R_2+L_2 M R_3-M M_w R_3,\\\notag
x_{6,4} &= R_2 L_2^2-2 R_2 M^2+L_2 M_w R_2,\\\notag
x_{6,5} &= L_2^2-2 M^2+L_2 M_w,\\\notag
x_{6,6} &= -R_2 L_2^2+2 R_2 M^2-L_2 M_w R_2-R_3 L_2^2+2 R_3 M^2-L_2 M_w R_3,\\\notag
x_{6,7} &= 2 R_2 M^2-L_2 M_w R_2-R_2 M_w^2,\\ \notag
x_{6,8} &= 2 M^2-L_2 M_w-M_w^2,\\\notag
x_{6,9} &= -2 R_2 M^2+L_2 M_w R_2+R_2 M_w^2-2 R_3 M^2+L_2 M_w R_3+R_3 M_w^2,\\\notag
x_{6,10} &= -L_2 M R_2+M M_w R_2,\\\notag
x_{6,11} &= -L_2 M+M M_w,\\\notag
x_{6,12} &= L_2 M R_2-M M_w R_2+L_2 M R_3-M M_w R_3,\\\notag
% \end{align}
% %
% \begin{align*}
% \notag
x_{9,1} &= -L_2 M R_2 + M M_w R_2,\\\notag
x_{9,2} &= -L_2  M + M  M_w,\\\notag
x_{9,3} &= L_2  M  R_2 - M  M_w  R_2 + L_2  M  R_3  - M  M_w  R_3 ,\\\notag
x_{9,4} &= 2  M^2  R_2 - L_2 M_w  R_2 - M_w^2  R_2,\\\notag
x_{9,5} &= 2  M^2 - L_2  M_w - M_w^2,\\\notag
x_{9,6} &= -2  M^2  R_2+ L_2  M_w  R_2 + M_w^2  R_2 - 2  M^2  R_3 + L_2  M_w  R_3 + M_w^2  R_3,\\\notag
x_{9,7} &=L_2^2  R_2 - 2  M^2  R_2 + L_2  M_w  R_2,\\\notag
x_{9,8} &= L_2^2 - 2  M^2 + L_2  M_w,\\\notag
x_{9,9} &= -L_2^2  R_2 + 2  M^2  R_2 - L_2  M_w  R_2 - L_2^2  R_3 + 2  M^2  R_3- L_2  M_w  R_3,\\\notag
x_{9,10} &= -L_2  M  R_2 + M M_w R_2,\\\notag
x_{9,11} &= -L_2  M + M  M_w,\\\notag
x_{9,12} &= L_2 M R_2 - M  M_w  R_2 + L_2  M  R_3 - M  M_w  R_3,\\\notag
% \end{align}
% \end{figure*}
% \begin{figure*}[!ht]
% \begin{align}
% \notag
x_{12,1} &= -2  M^2  R_2+ L_2  M_w  R_2 +M_w^2  R_2,\\ \notag
x_{12,2} &= -2  M^2 + L_2  M_w + M_w^2,\\\notag
x_{12,3} &= 2  M^2  R_2 - L_2  M_w  R_2 - M_w^2  R_2 + 2  M^2  R_3 - L_2  M_w  R_3 - M_w^2  R_3,\\\notag
x_{12,4} &= L_2  M  R_2 - M  M_w  R_2,\\\notag
x_{12,5} &= L_2  M - M  M_w,\\\notag
x_{12,6} &= -L_2  M  R_2 + M  M_w  R_2 - L_2  M  R_3 + M  M_w  R_3,\\\notag
x_{12,7} &= L_2  M  R_2 - M  M_w  R_2,\\\notag
x_{12,8} &= L_2  M - M  M_w,\\\notag
x_{12,9} &= -L_2  M  R_2 + M  M_w  R_2 - L_2  M  R_3 + M  M_w  R_3,\\\notag
x_{12,10} &= -L_2^2  R_2 + 2  M^2  R_2 - L_2  M_w  R_2,\\\notag
x_{12,11} &= -L_2^2 + 2  M^2 - L_2  M_w,\\\notag
x_{12,12} &= L_2^2  R_2 - 2  M^2  R_2 + L_2  M_w  R_2 + L_2^2  R_3 - 2 M^2  R_3 + L_2  M_w  R_3.\\\notag
\end{align}
}
\medmuskip=4mu
\thinmuskip=4mu
\thickmuskip=4mu

\normalsize
\section{Linear Model of Flux/Voltage Circuits with Plasma}
\label{Appendix:analytic-plasma-model}

When a plasma is present in the confinement volume, we model the interaction between the plasma and the injector circuits as an additional inductor coupling to the voltage and flux coils. We assume that the coupling between the plasma and voltage or flux coil is the same. This mutual inductance is referred to as $M_p$. This value is computed as
\begin{equation}
    M_p = K_i\sqrt{M_0L_2}.
\end{equation}
$M_0$ is the self-inductance of the plasma, and we take $K_i = 0.1$. These values can be tuned to obtain different behavior for the injector circuits, and the values used here were based on both measurement of the plasma self-inductance, and intuition about the experiment. \par
The structure of the model with plasma, $\mathbf{A}_{p}$, is very similar to $\mathbf{A}_{v}$. The equations for the first and second states, that being the current through the series coil, and the voltage across the capacitor, are the same, however the equations for the current through the flux or voltage coils will change. The structure of this matrix will also change from being in $\mathbb{R}^{12 \times 12}$ to $\mathbb{R}^{13 \times 13}$, as current through the coil representing the spheromak is taken to be a state variable. Again, all of the terms in rows three, six, nine, and twelve will have a prefactor, in this case,
\medmuskip=-2mu
\thinmuskip=-2mu
\thickmuskip=-2mu
\begin{align}
\label{eq:K_plasma_model}
    K = \frac{1}{(L_2 - M_w)(L_2 - 2M + M_w)(L_{p}(2M + M_w + L_2) - 4M_{p}^2}.
\end{align}
\medmuskip=4mu
\thinmuskip=4mu
\thickmuskip=4mu
In addition, there is another prefactor for the equations that  determines the current through the coil that represents the plasma: \begin{align}
\label{eq:K_plasma}K_{plasma} = \frac{1}{L_p(2M + M_w + L_2) - 4M_p^2}.
\end{align}
The coefficients of $\mathbf{A}_p$ are as follows:

% \begin{figure*}
\medmuskip=-3mu
\thinmuskip=-3mu
\thickmuskip=-3mu
\allowdisplaybreaks
\scriptsize{
\begin{align}
    x_{3,1} &=  -2L_{p}M^2R_{2} + 4MM_{p}^2R_{2} + L_{2}^2L_{p}R_{2} - M_{p}^2M_{w}R_2 - 3L_{2}M_{p}^2R_2 + L_{2}L_{p}M_{w}R_{2} ,\\\notag
    x_{3,2} &= -2L_{p}M^2 + 4MM_{p} + L_{2}^2L_{p} - M_{p}^2M_{w} - 3L_{2}M_{p}^2 + L_{2}L_{p}M_{w}, \\ \notag
    x_{3,3} &= (R_{2} + R_{3}) (2L_{p}M^2 - 4MM_{p}^2 - L_{2}^2L_{p} + M_{p}^2M_{w} + 3L_{2}M_{p}^2 - L_{2}L_{p}M_{w} ) ,\\\notag
    x_{3,4} &= L_{p}MM_{w}R_{2} - M_{p}^2M_{w}R_{2} + L_{2}M_{p}^2R_{2} - L_{2}L_{p}MR_{2} ,\\\notag
    x_{3,5} &= (R_{3} + R_{2})(-L_{p}MM_{w} + M_{p}^2M_{w} - L_{2}M_{p}^2 + L_{2}L_{p}M) ,\\\notag
    x_{3,6} &= -L_2 M R_2+M M_w R_2-L_2 M R_3+M M_w R_3,\\\notag
    x_{3,7} &= L_{p}MM_{w}R_{2} - M_{p}^2M_{w}R_{2} + L_{2}M_{p}^2R_{2} - L_{2}L_{p}MR_{2} ,\\\notag
    x_{3,8} &= L_{p}MM_{w} - M_{p}^2M_{w} + L_{w}M_{p}^2 - L_{p}ML_{2} ,\\\notag
    x_{3,9} &= (R_{2} + R_{3})(-L_{p}MM_{w} + M_{p}^2M_{w} - L_{2}M_{p}^2 + L_{p}ML_{2}) ,\\\notag
    x_{3,10} &= -L_{p}M_{w}^2R_{2} + 2L_{p}M^2R_{2} - 4MM_{p}^2R_{2} - 3M_{p}^2M_{w}R_{2} + L_{2}M_{p}^2R_{2} - L_{2}L_{p}M_{w}R_{2} ,\\\notag
    x_{3,11} &= -L_{p}M_{w}^2 + 2L_{p}M^2 - 4MM_{p}^2 + 3M_{p}^2M_{w} + L_{2}M_{p}^2 + L_{2}L_{p}M_{w},\\\notag
    x_{3,12} &= (R_{3} + R_{2})(L_{p}M_{w}^2 - 2L_{p}M^2 + 4MM_{p}^2 - 3M_{p}^2M_{w}-L_{2}M_{p}^2 + L_{2}L_{p}M_{w}), \\\notag
    x_{3,13} &= 2MM_{p}M_{w}R_{p} + L_{2}^2M_{p}R_{p} - M_{p}M_{w}^2R_{p} - 2L_{2}MM_{p}R_p \\\notag
% \end{align}
% \begin{align}
% \notag
x_{6,1} &= L_{2}L_{p}MR_{2} - L_{2}M_{p}^2R_{2} - L_{p}MM_{w}R_{2} + M_{p}^2M_{w}R_{2} ,\\\notag
x_{6,2} &= L_2 L_p M - L_2 M_p^2 - L_p M M_w  + M_p^2 M_w  ,\\\notag
x_{6,3} &= -L_2 L_p M R_2  + L_2 M_p^2 R_2  + L_p M M_w R_2  - M_p^2 M_w R_2  - L_2 L_p M R_3  + L_2 M_p^2 R_3  + L_p M M_w R_3  - M_p^2 M_w R_3  ,\\\notag
x_{6,4} &= -L_2^2 L_p R_2  + 2 L_p M^2 R_2  + 3 L_2 M_p^2 R_2 - 4 M M_p^2 +
R_2 - L_2 L_p M_w R_2 x_4 + M_p^2 M_w R_2  ,\\\notag
x_{6,5} &= -L_2^2 L_p + 2 L_p M^2 + 3 L_2 M_p^2 - 4 M M_p^2 - L_2 L_p M_w + M_p^2 M_w,\\\notag
x_{6,6} &= L_2^2 L_p R_2 - 2 L_p M^2 R_2 - 3 L_2 M_p^2 R_2 + 4 M M_p^2 R_2 + L_2 L_p + M_w R_2 - M_p^2 M_w R_2 + L_2^2 L_p R_3 - 2 L_p M^2 R_3 \\\notag& - 3 L_2 M_p^2 R_3 + 4 M M_p^2 R_3 + L_2 L_p M_w R_3 - M_p^2 M_w R_3,\\\notag
x_{6,7} &= -2 L_p M^2 R_2 - L_2 M_p^2 R_2 + 4 M M_p^2 R_2 + L_2 L_p M_w R_2 - 3 M_p^2 M_w R_2 + L_p M_w^2 R_2,\\ \notag
x_{6,8} &= -2 L_p M^2 - L_2 M_p^2 + 4 M M_p^2 + L_2 L_p M_w - 3 M_p^2 M_w + L_p M_w^2 ,\\\notag
x_{6,9} &=  L_p M^2 R_2 + L_2 M_p^2 R_2  - 4 M M_p^2 R_2 - L_2 L_p M_w R_2 + 3 M_p^2\\\notag & M_w R_2 - L_p M_w^2 R_2 + 2 L_p M^2 R_3 + L_2 M_p^2 R_3 - 
4 M M_p^2 R_3 - L_2 L_p M_w R_3 +
3 M_p^2 M_w R_3 - L_p M_w^2 R_3,\\\notag
x_{6,10} &= L_2 L_p M R_2 - L_2 M_p^2 R_2  - L_p M M_w R_2  + M_p^2 M_w R_2  ,\\\notag
x_{6,11} &= L_2 L_p M  - L_2 M_p^2  - L_p M M_w  + M_p^2 M_w  ,\\\notag
x_{6,12} &= -L_2 L_p M R_2 + L_2 M_p^2 R_2 + L_p M M_w R_2 - M_p^2 M_w R_2 -L_2 L_p M R_3  + L_2 M_p^2 R_3 + L_p M M_w R_3 - M_p^2 M_w R_3 , \\\notag
x_{6,13} &= -L_2^2 M_p R_p + 2 L_2 M M_p R_p - 2 M M_p M_w R_p + M_p M_w^2 R_p,\\\notag
% \end{align}
%
% \begin{align}
% \notag
x_{9,1} &= L_2 L_p M R_2  - L_2 M_p^2 R_2 - L_p M M_w R_2 + M_p^2 M_w R_2 ,\\\notag
x_{9,2} &= L_2 L_p M  - L_2 M_p^2  - L_p M M_w  + M_p^2 M_w ,\\\notag
x_{9,3} &= -L_2 L_p M R_2  + L_2 M_p^2 R_2  + L_p M M_w R_2 - M_p^2 M_w R_2  - L_2 L_p M R_3 + L_2 M_p^2 R_3 x_3 + L_p M M_w R_3  - M_p^2 M_w R_3 ,\\\notag
x_{9,4} &= -2 L_p M^2 R_2 - L_2 M_p^2 R_2 + 4 M M_p^2 R_2 + L_2 L_p M_w R_2 - 
 3 M_p^2 M_w R_2  + L_p M_w^2 R_2 ,\\\notag
x_{9,5} &= -2 L_p M^2 - L_2 M_p^2 + 4 M M+p^2 + L_2 L_p M_w - 3 M_p^2 M_w +
  L_p M_w^2 ,\\\notag
x_{9,6} &= 2 L_p M^2 R_2 + L_2 M_p^2 R_2 - 4 M M_p^2 R_2 - L_2 L_p M_w R_2 + \\\notag
 & 3 M_p^2 M_w R_2 - L_p M_w^2 R_2 + 2 L_p M^2 R_3 + L_2 M_p^2 R_3 - 4 M M_p^2 R_3 - L_2 L_p M_w R_3 + 3 M_p^2 M_w R_3  - L_p M_w^2 R_3,\\\notag
x_{9,7} &=-L_2^2 L_p R_2 + 2 L_p M^2 R_2 + 3 L_2 M_p^2 R_2 - 4 M M_p^2 R_2 - 
 L_2 L_p M_w R_2 + M_p^2 M_w R_2  ,\\\notag
x_{9,8} &= -L_2^2 L_p + 2 L_p M^2  + 3 L_2 M_p^2 - 4 M M_p^2 - L_2 L_p M_w +
  M_p^2 M_w ,\\\notag
x_{9,9} &= L_2^2 L_p R_2  - 2 L_p M^2 R_2 - 3 L_2 M_p^2 R_2 + 4 M M_p^2 R_2 + \\\notag
& L_2 L_p M_w R_2 - M_p^2 M_w R_2  + L_2^2 L_p R_3  - 2 L_p M^2 R_3 - 3 L_2 M_p^2 R_3 + 4 M M_p^2 R_3 + L_2 L_p M_w R_3 - M_p^2 M_w R_3 ,\\\notag
x_{9,10} &= L_2 L_p M R_2  - L_2 M_p^2 R_2  - L_p M M_w R_2  + M_p^2 M_w R_2 ,\\\notag
x_{9,11} &= L_2 L_p M  - L_2 M_p^2 - L_p M M_w + M_p^2 M_w ,\\\notag
x_{9,12} &= -L_2 L_p M R_2 + L_2 M_p^2 R_2  + L_p M M_w R_2 - M_p^2 M_w R_2 - L_2 L_p M R_3  + L_2 M_p^2 R_3  + L_p M M_w R_3  - M_p^2 M_w R_3 ,\\\notag
 x_{9,13} &= -L_2^2 M_p R_p + 2 L_2 M M_p R_p - 2 M M_p M_w R_p + M_p M_w^2 R_p \\\notag
% \end{align}
% \end{figure*}
% \begin{figure*}[!ht]
% \begin{align}
% \notag
x_{12,1} &= 2 L_p M^2 R_2 + L_2 M_p^2 R_2 - 4 M M_p^2 R_2 - L_2 L_p M_w R_2  + 
 3 M_p^2 M_w R_2 - L_p M_w^2 R_2 ,\\ \notag
x_{12,2} &= 2 L_p M^2 + L_2 M_p^2  - 4 M M_p^2 - L_2 L_p M_w + 3 M_p^2 M_w - 
 L_p M_w^2 ,\\\notag
x_{12,3} &= -2 L_p M^2 R_2- L_2 M_p^2 R_2 + 4 M M_p^2 R_2 +L_2 L_p M_w R_2 - 3 M_p^2 M_w R_2 + L_p M_w^2 R_2 -\\\notag&
2 L_p M^2 R_3 - L_2 M_p^2 R_3  + 
 4 M M_p^2 R_3 + L_2 L_p M_w R_3 - 3 M_p^2 M_w R_3 + L_p M_w^2 R_3 ,\\\notag
x_{12,4} &= -L_2 L_p M R_2 + L_2 M_p^2 R_2 + L_p M M_w R_2 - M_p^2 M_w R_2 ,\\\notag
x_{12,5} &= -L_2 L_p M + L_2 M_p^2 + L_p M M_w - M_p^2 M_w  ,\\\notag
x_{12,6} &= L_2 L_p M R_2 - L_2 M_p^2 R_2 - L_p M M_w R_2 + M_p^2 M_w R_2 + L_2 L_p M R_3- L_2 M_p^2 R_3 - L_p M M_w R_3 + M_p^2 M_w R_3,\\\notag
x_{12,7} &= -L_2 L_p M R_2 + L_2 M_p^2 R_2 + L_p M M_w R_2 - M_p^2 M_w R_2 ,\\\notag
x_{12,8} &= -L_2 L_p M + L_2 M_p^2 + L_p M M_w - M_p^2 M_w,\\\notag
x_{12,9} &= L2 L_p M R_2 - L_2 M_p^2 R_2 - L_p M M_w R_2 + M_p^2 M_w R_2+ L_2 L_p M R_3 - L_2 M_p^2 R_3 - L_p M M_w R_3+ M_p^2 M_w R_3,\\\notag
x_{12,10} &= L_2^2 L_p R_2 - 2 L_p M^2 R_2 - 3 L_2 M_p^2 R_2 + 
 4 M M_p^2 R_2 + L_2 L_p M_w R_2- M_p^2 M_w R_2 ,\\\notag
x_{12,11} &= L_2^2 L_p - 2 L_p M^2 - 3 L_2 M_p^2 + 4 M M_p^2 + 
 L_2 L_p M_w- M_p^2 M_w,\\\notag
x_{12,12} &= -L_2^2 L_p R_2 + 2 L_p M^2 R_2 + 3 L_2 M_p^2 R_2 - 4 M M_p^2 R_2- L_2 L_p M_w R_2 + M_p^2 M_w R_2 - \\\notag
 & L_2^2 L_p R3 + 2 L_p M^2 R_3  + 3 L_2 M_p^2 R_3 - 4 M M_p^2 R_3 - L_2 L_p M_w R_3+ M_p^2 M_w R_3.\\\notag
 x_{12,13} &= L_2^2 M_p R_p - 2 L_2 M M_p R_p + 2 M M_p M_w R_p - M_p M_w^2 R_p, \\ \notag
    x_{13a} &=  M_p R_2 ,\\\notag
    x_{13b} &= M_p,\\\notag
    x_{13c} &= -M_p R+2 - M_p R_3,\\\notag
    x_{13d} &= M_p R_2,\\\notag
    x_{13e} &= M_p,\\\notag
    x_{13f} &= -M_p R_2  - M_p R_3 ,\\\notag
    x_{13g} &=  M_p R_2,\\\notag
    x_{13h} &=  M_p,\\\notag
    x_{13i} &= -M_p R_2 - M_p R_3,\\\notag
    x_{13j} &= M_p R_2,\\\notag
    x_{13k} &= M_p,\\\notag
    x_{13l} &= M_p R_2 - M_p R_3\\    \notag
    x_{13m} &= L_2 R_p + 2 M R_p + M_w R_p. \notag
\end{align}

}

\newpage
\normalsize
\section{Actuation Matrix for the Injector Circuits}
\label{Appendix:actuation-matrix} 
The actuation matrix for this system is solely a function of the inductance of the series inductor in the circuit shown in Fig. \ref{fig:circuit_topology}. There is a factor of $1/L_1$ that repeats in the first, fourth, seventh, and tenth rows of the matrix. Each column in this matrix represents one of the four independent power supplies. The factor of $1/L_1$ shifts by one column corresponding to the actuation provided to each circuit. 

\begin{align}\notag
   \boldsymbol{B}  &= \begin{bmatrix}
        1/L_1 \!\!\!\!\!\!\!\!\!\!\!\! &  0 \!\!\!\!\!\!\!\!\!\!\!\! & 0 \!\!\!\!\!\!\!\!\!\!\!\! & 0  \\
        0 \!\!\!\!\!\!\!\!\!\!\!\!& 0 \!\!\!\!\!\!\!\!\!\!\!\! & 0 \!\!\!\!\!\!\!\!\!\!\!\! & 0 \\
        0 \!\!\!\!\!\!\!\!\!\!\!\! & 0 \!\!\!\!\!\!\!\!\!\!\!\! & 0 \!\!\!\!\!\!\!\!\!\!\!\! & 0 \\
        0 \!\!\!\!\!\!\!\!\!\!\!\! & 1/L_1 \!\!\!\!\!\!\!\!\!\!\!\! & 0 \!\!\!\!\!\!\!\!\!\!\!\! & 0 \\
        \!\!\!\!\!\!\!\!\!\!\!\! & \vdots & & \\
        0 \!\!\!\!\!\!\!\!\!\!\!\!& 0 \!\!\!\!\!\!\!\!\!\!\!\! & 0 \!\!\!\!\!\!\!\!\!\!\!\! & 1/L_1 \\
        \!\!\!\!\!\!\!\!\!\!\!\! & \vdots & & \\  0 \!\!\!\!\!\!\!\!\!\!\!\!& 0 \!\!\!\!\!\!\!\!\!\!\!\! & 0 \!\!\!\!\!\!\!\!\!\!\!\! & 0 \\
    \end{bmatrix}
\end{align}

% \nocite{*}
\clearpage

\bibliography{references}% Produces the bibliography via BibTeX.

\end{document}